%% file: MSPESII_rev4.tex
\documentclass[preprint2]{aastex}
\usepackage{longtable}
\usepackage{lscape}
\def\etal{\it et~al.}

\shorttitle{The phenomenon of Drifting Subpulses}
\shortauthors{Basu \etal}

\begin{document}

\title{Meterwavelength Single-pulse Polarimetric Emission Survey II: The phenomenon of Drifting Subpulses}

\author{Rahul Basu\altaffilmark{1}, Dipanjan Mitra\altaffilmark{2,3,1}, George I. Melikidze\altaffilmark{1,4}, Krzysztof Maciesiak\altaffilmark{1}, Anna Skrzypczak\altaffilmark{1}, Andrzej Szary\altaffilmark{5,1}} 

\altaffiltext{1}{Janusz Gil Institute of Astronomy, University of Zielona G\'ora, ul. Szafrana 2, 65-516 Zielona G\'ora, Poland}
\altaffiltext{2}{National Centre for Radio Astrophysics, Ganeshkhind, Pune 411 007, India}
\altaffiltext{3}{Physics Department, University of Vermont, Burlington VT 05405}
\altaffiltext{4}{Abastumani Astrophysical Observatory, Ilia State University, 3-5 Cholokashvili Ave., Tbilisi, 0160, Georgia}
\altaffiltext{5}{ASTRON, the Netherlands Institute for Radio Astronomy, Postbus 2, 7990 AA, Dwingeloo, The Netherlands}
\email{rahul@astro.ia.uz.zgora.pl}

\begin{abstract}
\noindent
A large sample of pulsars was observed as part of the Meterwavelength
Single-pulse Polarimetric Emission Survey.  We carried out a detailed
fluctuation spectral analysis which revealed periodic features in 46\%
pulsars including 22 pulsars where drifting characteristics were
reported for the first time.  The pulsar population can be categorized
into three distinct groups, pulsars which show systematic drift motion
within the pulse window, the pulsars showing no systematic drift but
periodic amplitude fluctuation and pulsars with no periodic
variations.  We discovered the dependence of the drifting phenomenon
on the spin down energy loss ($\dot{E}$), with the three categories
occupying distinctly different regions along the $\dot{E}$ axis.  The
estimation of the drift periodicity ($P_3$) from the peak frequency in
the fluctuation spectra is ambiguous due to the aliasing effect.
However, using basic physical arguments we were able to determine
$P_3$ in pulsars showing systematic drift motion.  The estimated $P_3$
values in these pulsars were anti-correlated with $\dot{E}$ which
favoured the Partially Screened Gap model of Inner Acceleration Region
in pulsars.
\end{abstract}


\section{\large Introduction} \label{sec:intro}
\noindent
The pulsar radio emission consist of a sequence of highly periodic
pulses occupying a small fraction of the pulsar period.  The
individual pulses are made up of one or more subpulses which, in some
cases, exhibit systematic variation in position or intensity or both.
This phenomenon is clearly seen in a pulse stack, which is an
alternative representation of the pulsar data where consecutive pulsar
periods are arranged on top of each other in a two dimensional array,
with the horizontal axis along the pulse longitude and the vertical
axis representing increasing period number.  An impression resembling
drift bands is seen in the pulse stack, first observed by
\citet{dra68}, and is known as the phenomenon of drifting subpulses in
pulsars, hereafter simply referred to as drifting.

The drifting greatly varies in the pulsar population and is broadly
classified into two groups.  The first corresponds to the phase
modulated drifting where regular drift bands are seen.  The second
case is the amplitude modulated drifting where the subpulses are
localised in the pulse window and only show periodic intensity
variation.  Around 35\% of pulsars have been reported to show some
form of drifting \citep{wel06,wel07}.  There are two periodicity
associated with drifting, $P_2$ which measures the longitudinal
separation between adjacent drift bands and $P_3$ the interval between
the signal repeating at the same location \citep{bac70,bac73}.

The drifting is closely related to the physical processes responsible
for radio emission in pulsars.  
A force-free condition is believed to exist around the neutron star which 
introduces an electric field ($\mathcal{E}$) at a radial distance ${\mathbf r}$ 
from the neutron star, which is represented in the observer's frame of 
reference as:
\begin{equation}
\mathcal{E} = - \frac{1}{c} \left({\mathbf\Omega}\times{\mathbf r}\right)\times{\mathbf B}.
\end {equation} 
where ${\mathbf\Omega}$(=2$\pi/P$) is the angular velocity of the neutron star 
and ${\mathbf B}$ is the magnetic field.
This further requires the formation of a charge separated
magnetosphere with density $n_{\rm GJ} = {\mathbf\Omega} \cdot
{\mathbf B}/2\pi e c$ \citep{gol69}, co-rotating with the neutron star.
It should be noted that if the charge density goes below $n_{\rm GJ}$,
they no longer co-rotate with the neutron star, but lags behind the
co-rotation with motion still around the rotation axis.

The pulsar magnetosphere is divided into two regions, the closed field
line region and the open field line region, bounded by the light
cylinder where the co-rotation speed is equal to the speed of light
(R$_{LC}$ = $c/\Omega$).  The magnetosphere in the open field lines is
initially charge starved, and a supply of charges can come from the
neutron star surface or magnetic e$^-$e$^+$ pair production.  An
abundant supply of plasma ensures a flow of relativistic charged
particles along the open field lines \citep{spi11}.  The pulsar radio
emission is believed to originate as a result of the growth of
instabilities in this outflowing plasma at a height of about 500 km
above the neutron star surface \citep{mit02,kij03,krz09,mel00,gil04}.
The physical processes generating the relativistic plasma require the
formation of a non-stationary inner acceleration region (IAR) in the
immediate vicinity of the polar cap.  The prototype for the IAR is the
inner vacuum gap (IVG) first suggested by \citet{rud75}, hereafter
RS75.  They used magnetic fields of magnitude $\sim10^{12}$G
calculated from slowdown rates corresponding to the dipolar component.
However, significantly smaller values for the radius of curvature was
used to determine the Lorenz factors of the relativistic charged
particles in the gap implying non-dipolar components of magnetic
field.  The electric potential difference in the gap can accelerate 
positron/electron with energies $\sim$ 10$^{12}$ eV.
RS75 suggested that the gap breaks down as a result of pair production in
high magnetic fields and subsequent acceleration in the electric
fields which give rise to several equispaced sparking discharges.  The
fully formed spark in the gap results in a non-stationary flow of
plasma column and radio emission generated in this column is observed
as a subpulse.  It should be noted that the sparks, i.e. the regions
of IAR where the local charge density differs from $n_{\rm GJ}$, do
not corotate with the pulsar.  This lack of corotation in the sparks
is manifested as the phenomenon of subpulse drifting in pulsars.

In recent years two methodologies have been used to probe the IAR, the
X-ray emission from the polar caps and measuring the drift velocities
\citep{gil08}.  The thermal component of the X-ray emission revealed
that the surface temperatures were lower than expected from the IVG of
RS75.  In certain pulsars the estimated drift velocities were less
than the predicted RS75 values (see \citealt{des99,gil03a} and
references therein).  This motivated \citet{gil03a} to propose the
partially screened gap (PSG) model for the IAR.

In this paper, Meterwavelength Single-pulse Polarimetric Emission
Survey-II (MSPESII), we have conducted an extensive study of drifting
in the sample of pulsars from MSPES, the details of which are
presented in section \ref{sec:obs} (observation and analysis),
\ref{sec:res} (primary results) and \ref{sec:cor} (drifting
properties).  The inherent ambiguity in determining $P_3$ from the
measured drifting feature due to the aliasing effect is discussed in
section \ref{sec:alias}.  In section \ref{sec:model} we have used
basic physical arguments to determine $P_3$ in certain pulsars with
interesting implications for the PSG model of the IAR.

\section{\large Observation and Analysis} \label{sec:obs}
\noindent
The details of the observing procedure of MSPES, using the Giant
Meterwave Radio Telescope, was reported in \citet{mit15}, see table 1
therein.  A total of 123 pulsars were observed in this survey with 118
pulsars at 618 MHz, 105 pulsars at 333 MHz and 100 pulsars at both
these frequencies.  The data were recorded in the full polarization
mode, but we used only the total intensity single pulses for these
studies\footnote{The pulsar J1703$-$3241 was observed on two occasions
  at 333 MHz, 2 March 2014 and 5 May 2014. MSPES uses data from 5 May
  due to proper polarization setting, we used 2 March data for MSPESII
  as the total intensity single pulses were found to be stronger.}.
We used the fluctuation spectral analysis for determining the periodic
features as described below.  Some additional pre-processing of the
data were also carried out as reported in appendix
\ref{sec:data_process}.

\subsection{\large Fluctuation Spectral Analysis} \label{sec:fluc_spec}
\begin{figure*}
\begin{center}
\includegraphics[angle=0,scale=0.70]{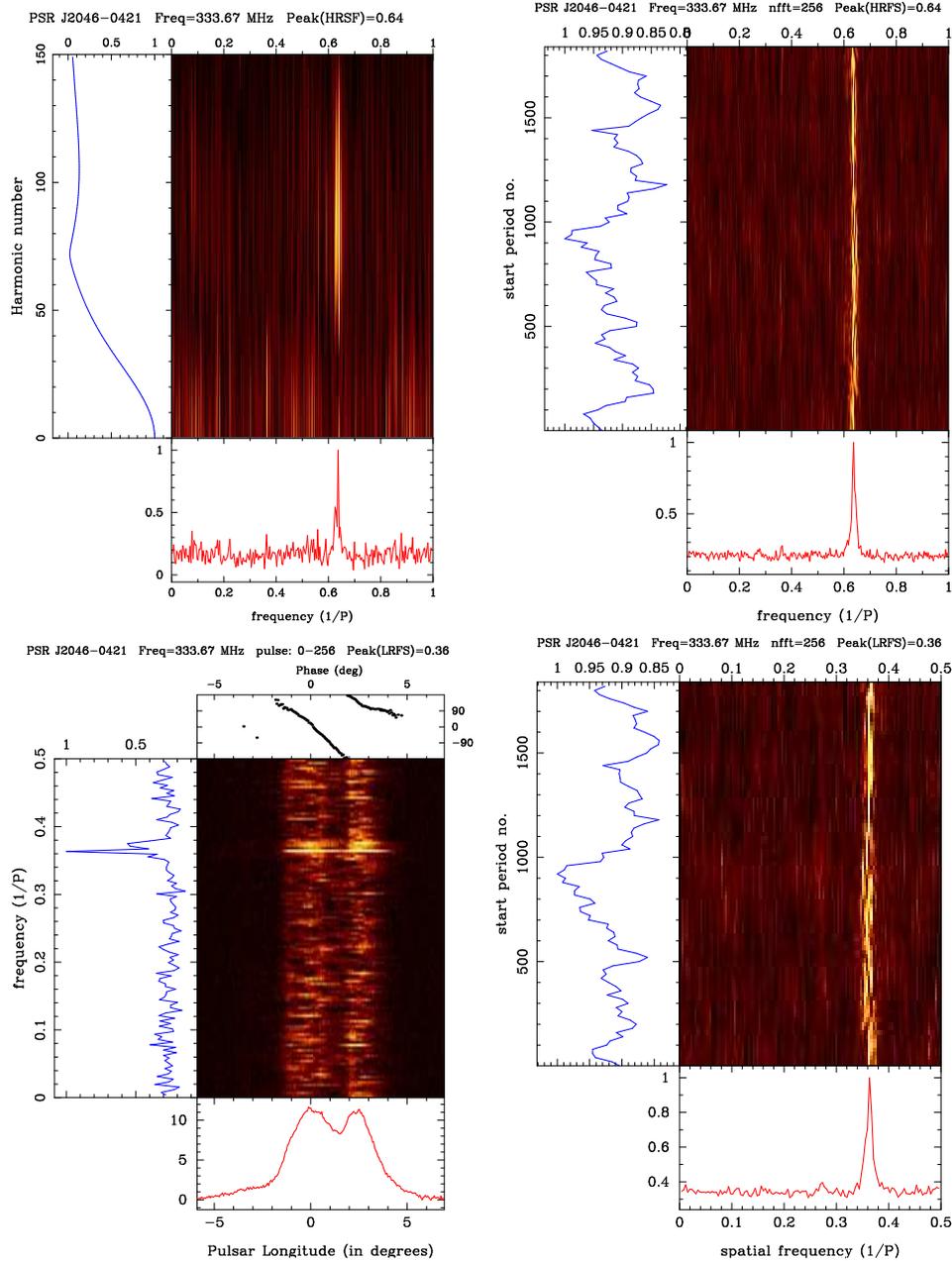} 
\end{center}
\caption{The fluctuation spectral analysis for the pulsar
  J2046$-$0421.  The LRFS (bottom left) and the HRFS (top left) is
  shown for the pulse sequence 0-256.  The LRFS also shows the phase
  variation across the pulse window at the peak amplitude.  The time
  variation of the LRFS (bottom right) and the HRFS (top right) was
  used to determine the peaks.  The LRFS in this example show a narrow
  peak at $f_p$ = 0.363 cycle/$P$ and the phase at the peak vary with
  a negative slope across the pulse window.  The corresponding peak in
  the HRFS is shifted to {\it\={f}$_p$} = 0.637 cycle/$P$, (1-$f_p$),
  which is another indication of the negative slope of phase
  variation.
\label{fig_flsp}}
\end{figure*}

\noindent
The most widely used technique for studying drifting is the
Fluctuation Spectral analysis using the method of Fourier Transforms,
where the peaks in the spectrum corresponds to the frequencies of
drifting.  There are three principal types reported in the literature,
the Longitude Resolved Fluctuation Spectra \citep[LRFS,][]{bac70}, the
Harmonic Resolved Fluctuation Spectra \citep[HRFS,][]{des01} and the
2-Dimensional Fluctuation Spectra \citep[2DFS,][]{edw02}.

The LRFS involve selecting an appropriate number of consecutive
pulses, $n_s$, and performing Discrete Fourier Transform (DFT) along
each longitude in the pulse stack.  The resultant complex fourier
transform can be separated into two parts, the amplitude with peaks
($f_p$) representing the frequencies of periodic fluctuation (in units
of cycles/$P$), and the phase at $f_p$ giving us a sense of subpulse
variation across the pulse window.

The HRFS provides an alternate method of estimating the fluctuations
in the pulsar signal and is useful in determining the nature of phase
variation especially in less bright pulsars where the phase variation
in the LRFS is not easily measured.  In this technique one single DFT
is carried out over the entire time series data, after padding the
off-pulse region with zeros, i.e. a $n_s \times n_p$ DFT is carried
out ($n_p$ is the number of phase bins in every pulse period).  The
resultant one dimensional fourier transform is arranged as a function
of the harmonic number (successive sections of 1/$P$ cycles) to form
the HRFS.  In contrast to the LRFS, where $f_p$ is mapped within the
frequency range 0 -- 0.5 cycles/$P$, the peaks in HRFS are mapped
between 0 -- 1.0 cycles/$P$.  The extra frequency space (0.5 -- 1.0
cycles/$P$) in the HRFS provides direct realizations of the phase
behaviour seen in the LRFS.

The 2DFS is an extension of the LRFS where additional Fourier
Transform is performed along each horizontal axis of the LRFS.  The
adjacent drift bands are repeated at any longitude ($\phi$) with a
periodicity $P_2$, implying a phase change of $\phi P/P_2$ across the
longitudes.  The estimation of peaks along the horizontal axis of 2DFS
gives a direct estimate of $P_2$.  Any $f_p$ in the $-$0.5 -- 0
cycles/$P$ frequency range of 2DFS is mapped in 0.5 -- 1.0 cycles/$P$
region of the HRFS.

We were primarily interested in estimating $f_p$ using the fluctuation
spectral technique (we used separate tools for estimating $P_2$ as
explained below) and therefore used the LRFS and HRFS for our
analysis.  The $n_s$ needed to carry out the LRFS/HRFS is ambiguous,
but we found 256 pulses to be optimal for our studies.  This number
was high enough to obtain sufficient resolution in the frequency
domain (0.004 cycle/$P$), yet it was not too high such that the power
in each frequency bin would be low for proper peak detection.  In
order to account for intrinsic intensity variation every 256 pulses
used for fluctuation spectral analysis were folded to form a profile
and all the 256 pulses subsequently normalised by the profile peak
before the fluctuation spectral analysis.  We searched for temporal
variation in drifting using both LRFS and HRFS.  The fluctuation
spectra were determined as a function of time after shifting the start
period by about 10 pulses and performing the DFT for each data set (a
similar technique was also proposed by \citealt{ser09} for the 2DFS).
The LRFS for each time realization was averaged along the longitude
while the HRFS was averaged along the harmonic number and depicted in
2-dimensional maps with the vertical axis as the start period.  The
time varying fluctuation spectra in both cases were finally averaged
across the time axis to determine the average frequency behaviour of
the drifting signal which was used to determine $f_p$.  The time
averaged fluctuation spectra were normalised by the maximum value in
spectra to make it independent of absolute flux.  An example of our
analysis is shown in figure \ref{fig_flsp}, with the LRFS and HRFS
(left panel, bottom and top respectively) as well as their time
variations (right panel).

\subsection{\large Determining $f_p$} \label{sec:P3measure}
\noindent
The time averaged LRFS and HRFS were used to determine the frequency
peaks of fluctuation as described below:\\ 
1. The fluctuation spectra
was divided into five sections and a mean and rms was determined for
each.  The section with the minimum rms was identified as the baseline
with its mean ($\mu_f$) and rms ($\sigma_f$) used as baseline
characteristics.\\ 
2. All regions in the spectra with at least 3
consecutive points in excess of $\mu_f$+5$\sigma_f$ were identified as
potential structures.  In case of wider structures the points going
below the cutoff level was also included provided they were bordered
by values in excess of the cutoff level.\\ 
3. The peak frequency was
determined as the centroid of the above identified regions
\citep{wel06}:
\begin{equation}
f_p = \frac{\sum V_f\times f}{\sum V_f},
\end{equation}
here $V_f$ is the fluctuation spectra value at frequency $f$.\\ 
4. The Full Width at Half Maximum ($FWHM$) for each peak, around the
region identified for peak measurement, was calculated and used to
estimate the rms level (gaussian approximation):
\begin{equation}
\sigma_p = \frac{FWHM}{2\sqrt{2~\mbox{ln}2}} \approx \frac{FWHM}{2.355}
\end{equation}
5. The error in $f_p$ was estimated as \citep{kij97}:
\begin{equation}
\Delta f_p = \sigma_p\sqrt{1+\left(\frac{\sigma_f}{V_p-\mu_f}\right)^2}
\end{equation}
Here $V_p$ is the value of the fluctuation spectra at $f_p$.

\subsection{\large Determining $P_2$}
\begin{figure*}
\begin{center}
\includegraphics[angle=0,scale=0.70]{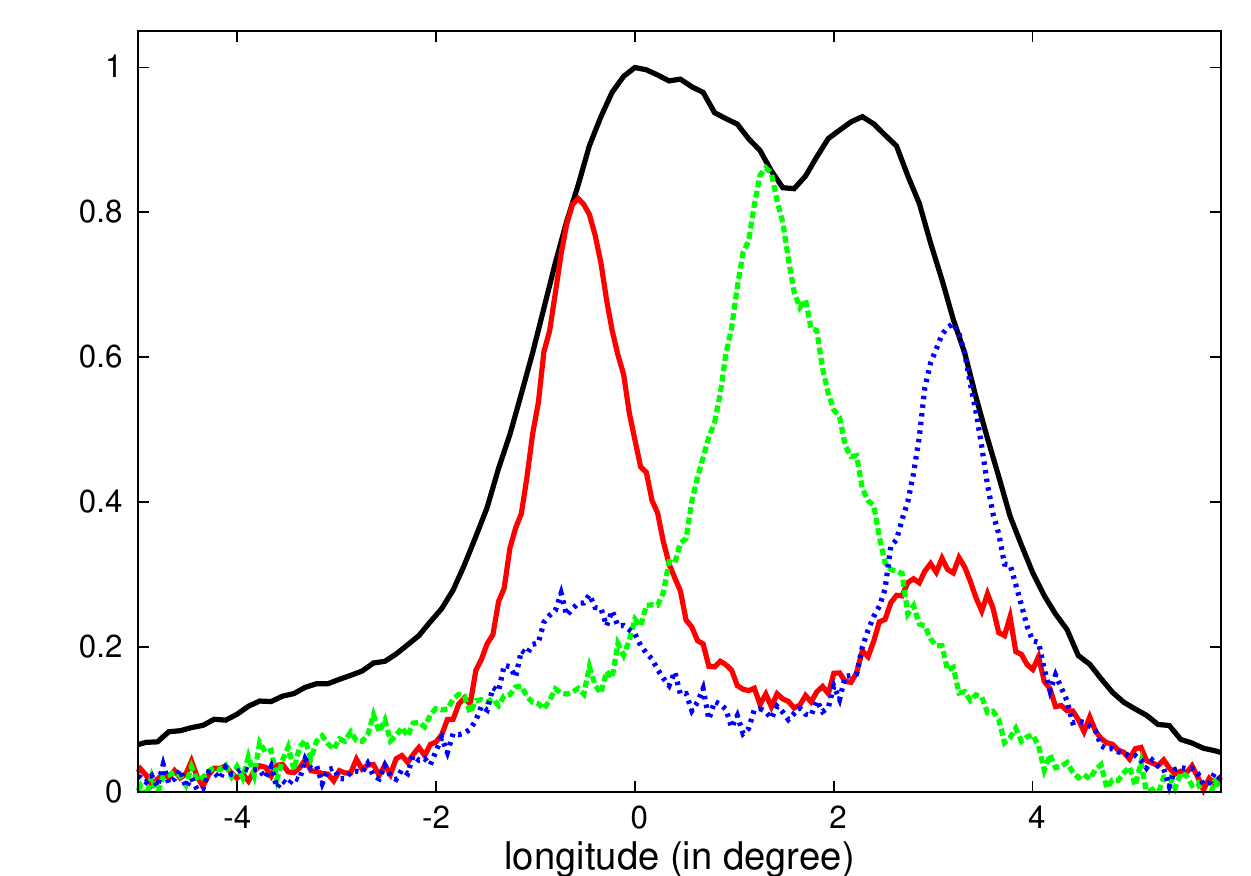} 
\caption{The figure shows the Peak folding method for estimating $P_2$
  in pulsar J2046$-$0421.  The single pulses with peaks within a
  narrow window (5 bins) are averaged to form profile with separated
  subpulses.  The window is shifted across the pulse longitudes from
  the leading to the trailing edge to form multiple profiles.  Three
  such examples are shown in the figure, the red profile with peak
  near the leading edge, the green profile with peak near the central
  region and the blue profile with peak near the trailing edge, with
  the average profile in black.  The red and blue profiles have well
  separated subpulses and can be used to determine the separation
  between subpulse peaks.  The estimated $P_2$ is the mean of all such
  separations that can be measured.  The estimated $P_2$ for PSR
  J2046$-$0421 is 3.37$\pm$0.06\degr.
\label{fig_P2}}
\end{center}
\end{figure*}

\noindent
The $P_2$ can be measured unambiguously if there are more than one
drift band seen in a given single pulse, making it viable in a small subset 
of pulsars.  A direct method is to measure separation between the subpulses, 
corresponding to adjacent drift bands, from strong single pulses.  An 
enhancement to this approach was devised in order to increase the signal to 
noise ratio.  The peaks in the pulse window for all significant single pulses 
($>$ 5$\sigma_N$, $\sigma_N$ the rms of off-pulse region) were determined.  A 
narrow window in the pulse longitude (3-5 bins) was selected and all the single
pulses with peak within this window were averaged to form a folded profile with
well separated subpulses.  The selected window was shifted continuously across 
the entire pulse window to generate multiple profiles.  The separation between 
the peaks of adjacent subpulses in each profile (if more than one peak was 
visible) was measured.  $P_2$ was determined after averaging the measured 
separation from all profiles.  The error in $P_2$ is given as $\Delta P_2$ =
$\Delta\phi\sqrt{1+(\sigma_P/V_P)^2}$, here $\Delta\phi$ is the resolution of 
profile, $\sigma_P$ the rms in the off pulse region of the average profile and 
$V_P$ the peak value of average profile.  In figure \ref{fig_P2} we show the 
result of these exercises for the pulsar J2046$-$0421.  Three peak-folded sub 
profiles are shown in the figure (red, green and blue profiles) as well as the 
full profile from all single pulses (in black).  As seen in the figure the 
profiles in red and blue has well separated subpulses and all such profiles 
were used for estimating $P_2$.  The green profile has just one detectable peak
and hence could not be used for these calculations.


\section{\large Results} \label{sec:res}
\begin{table*}
\small
\begin{center}
\caption{List of Pulsars with Drifting features
\label{drift_list}}
{\begin{tabular}{ccc|ccc|cc}
\multicolumn{3}{c}{Phase Modulation} & \multicolumn{3}{c}{Amplitude Modulation} & \multicolumn{2}{c}{No Detection} \\
\tableline
 New Detection & \multicolumn{2}{c}{Previously Reported} & New Detection & \multicolumn{2}{c}{Previously Reported} &  \\
\tableline
    PSR       &   PSR        &   Ref. &    PSR    &   PSR    & Ref. &   PSR    & Ref. \\
\tableline
 J0846$-$3533 & J0034$-$0721 & 1, 2, 3, 4, 5, 6 & J1034$-$3224$^*$ & J0758$-$1528 & 5 & J0820$-$4114 & 10, 11 \\
 J0959$-$4809 & J0151$-$0635 & 4, 5, 7 & J1116$-$4122 &  J0837+0610  & 4, 5, 15, 16 &  J0922+0638  & 4, 5 \\
 J1418$-$3921 & J0152$-$1637 & 4, 5 & J1328$-$4921 &  J1239+2453  & 4, 5, 19, 20, 21  &  J0953+0755  & 5, 17, 18 \\
 J1527$-$3931 &  J0304+1932  & 4, 5, 8, 9 & J1603$-$2531 & J1645$-$0317 & 4, 5, 22, 23 & J1607$-$0032 & 4, 5 \\
 J1555$-$3134 &  J0525+1115  & 4, 5 & J1604$-$4909 & J1733$-$2228 & 5 & J1820$-$0427 & 5 \\
 J1703$-$3241 & J0630$-$2834 & 4, 5 & J1625$-$4048 & J1735$-$0724 & 5 & J1847$-$0402 & 5 \\
 J1700$-$3312 & J0820$-$1350 & 4, 5, 12, 13, 14 & J1722$-$3207 &  J1740+1311  & 5, 24, 25 & J1849$-$0636 & 4, 5 \\
 J1816$-$2650 & J0944$-$1354 & 5 & J1733$-$3716 & J1842$-$0359 & 5 & J1913$-$0440 & 4 \\
  J1919+0134  & J1041$-$1942 & 4, 5 & J1741$-$3927 & J1848$-$0123 & 4, 26 & J1941$-$2602 & 4 \\
              & J1720$-$2933 & 4, 5 & J1748$-$1300 & J1900$-$2600 & 4, 5, 27 & J2346$-$0609 & 4 \\
              & J1741$-$0840 & 4, 5 & J1801$-$0357 &  J1909+1102  & 5 &              &  \\
              & J1822$-$2256 & 4, 5, 6 & J1801$-$2920 &  J1919+0021  & 4, 5 &              &  \\
              & J1901$-$0906 & 4, 5 & J2006$-$0807 &  J1932+1059  & 4, 5, 29, 30, 31 &              &  \\
              &  J1921+1948  & 5, 26, 28 &              &  J1946+1805  & 4, 5, 6, 32, 33 &              &  \\
              &  J1921+2153  & 4, 5, 19 &              & J2048$-$1616 & 4, 5, 22, 30, 34 &              &  \\
              & J2046$-$0421 & 4, 5 &              & J2330$-$2005 & 5 &              &  \\
              &  J2046+1540  & 4, 5 &              &              &   &              &  \\
              &  J2305+3100  & 4, 5, 35, 36 &              &              &   &              &  \\
              &  J2317+2149  & 5 &              &              &   &              &  \\
\tableline
\end{tabular}}
\tablenotetext{~}{$^*$ the phase change was indeterminate;\\ References: 1-\citealt{hug70}; 2-\citealt{viv97}; 3-\citealt{smi05}; 4-\citealt{wel06}; 5-\citealt{wel07}; 6-\citealt{ser09}; 7-\citealt{big85}; 8-\citealt{sch73}; 9-\citealt{bac75}; 10-\citealt{bha07}; 11-\citealt{bha09}; 12-\citealt{big87}; 13-\citealt{jan04}; 14-\citealt{lyn83}; 15-\citealt{sut70}; 16-\citealt{asg05}; 17-\citealt{bac70}; 18-\citealt{wol80}; 19-\citealt{pro86}; 20-\citealt{sro05}; 21-\citealt{maa14}; 22-\citealt{tay71}; 23-\citealt{tay75}; 24-\citealt{ran88}; 25-\citealt{for10}; 26-\citealt{han87}; 27-\citealt{mit08}; 28-\citealt{ran13}; 29-\citealt{ost77b}; 30-\citealt{now82}; 31-\citealt{bac73}; 32-\citealt{dei86}; 33-\citealt{klo10}; 34-\citealt{ost77a}; 35-\citealt{sie75}; 36-\citealt{red05}.}
\end{center}
\end{table*}

\noindent
The analysis techniques described in the previous section were implemented using software packages developed for this work and applied to the 123 pulsars in MSPES.
The pulsars showing drifting are listed in Table \ref{drift_list}.
In Table \ref{peak_mes} we report the detected peaks in time average LRFS and HRFS for each pulsar. 
All significant peaks along with $P_2$ (if applicable) are also listed in the table.

\subsection{Summary of Results}
\noindent
1. 57 pulsars has features with measurable peaks, i.e. 46\% of pulsars
in our sample show drifting.\\

\noindent
2. Drifting is seen for the first time in 22 pulsars and verified in
further 35 cases.  There are around 100 pulsars with drifting reported
in the literature \citep{gil06}.  Our present survey substantially
increases this population by 20\%.  The references for previous
studies of drifting in individual pulsars are listed in
table~\ref{drift_list}.\\

\noindent
3. We have detected drifting in 38 pulsars at 333 MHz, 44 pulsars at
618 MHz and 25 at both frequencies.\\

\noindent
4. $P_2$ is measured in 10 pulsars, i.e. around 18\% of the pulsars
with drifting.\\

\noindent
5. In 28 pulsars the periodic features show non zero phase variations,
9 of which are new detections. \\

\noindent
6. In 29 pulsars the periodic features are associated with no phase
variation, 13 of which are reported for the first time. \\

\noindent
7. There are 10 pulsars where drifting has been reported in the past
but our analysis failed to detect any such features.  In most cases
this could be traced to the presence of RFI or low level features
below our detection threshold.

\subsection{Classification of Drifting Features}
\subsubsection{The strength of Drifting feature} 
\noindent
The drifting was classified by \cite{wel06} in terms of the width of
the features, narrow widths ($< 0.05$ cycles/$P$) were labeled as
coherent drifters while wider features ($> 0.05$ cycles/$P$) were
called diffuse drifters.  This classification scheme do not include
any quantitative information about the relative strength of the
features.  We have devised an updated scheme for quantifying drifting
using the quantity $S$ which is defined as the ratio between the peak
height and effective width of the feature;
\begin{equation}
S = \frac{V_p - \mu_f}{FWHM}, 
\end{equation}
here $V_p$ is the measured value at $f_p$, $\mu_f$ the mean baseline
level and $FWHM$ is the Full Width at Half Maximum of the drifting
feature.  We determined $S_L$ and $S_H$ for all peaks measured from
the time averaged LRFS and HRFS (table \ref{peak_mes}).  The
normalisation of the fluctuation spectra (see section
\ref{sec:fluc_spec}) ensured that the estimated $S$ is independent of
the large scale intensity fluctuation as well as the absolute flux of
the pulsar.

\subsubsection{The Phase behaviour}\label{sec:phase}
\begin{figure*}
\begin{center}
\includegraphics[angle=0,scale=0.70]{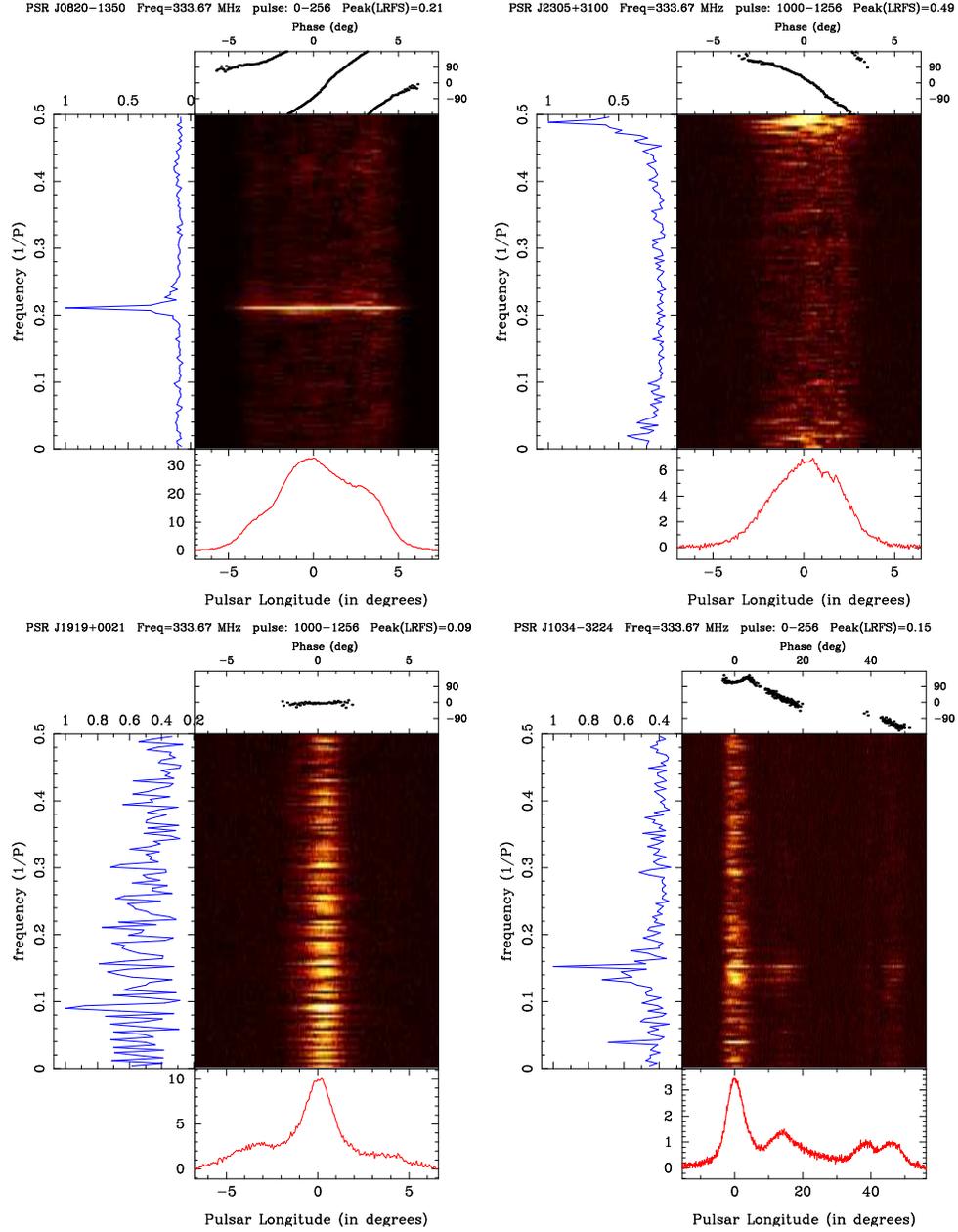} 
\end{center}
\caption{The figure shows the LRFS corresponding to the four different
  configurations of drifting based on the phase behaviour.  The pulsar
  J0820$-$1350 (top left panel) belongs to the class Negative Drifting
  (ND) where the phase show a positive slope as it changes across the
  pulse window.  The pulsar J2305$+$3100 (top right panel) is an
  example of Positive Drifting (PD) with the phase showing a negative
  slope from the leading to the trailing edge.  In case of pulsar
  J1919+0021 (bottom left panel) no phase change seen across the
  profile which is an example of Amplitude Modulated Drifting (AMD).
  The pulsar J1034$-$3224 (bottom right panel) show a complicated
  phase behaviour which cannot be classified into the three
  categories.
\label{fig_phasevar}}
\end{figure*}

\noindent
The drifting can be classified into different categories based on the
nature of subpulse motion across the pulse window.  This is reflected
in the phase behaviour seen at the peak frequency in LRFS, either
showing large phase variation or very little change in phase across
the pulse window.  Sometimes, the periodic features are too weak with
no phase measurements possible.  In these cases the HRFS can be used
to decipher the phase nature as it shows different behaviour for each
case.  Based on the phase behaviour we have divided the drifting
population into three groups which we describe below.\\

\noindent
{\bf Negative Drifting (ND)}: This category of pulsars exhibit phase
variation across the pulse window which show a positive slope from the
leading to the trailing edge of the profile.  This is exemplified by
the pulsar J0820$-$1350~in figure \ref{fig_phasevar}.  The ND is seen
in the pulse stack as subpulses shifting towards the leading edge of
the profile.  The HRFS has a unique response for the ND, with the
peaks in the LRFS having identical locations in the HRFS as well, i.e
the HRFS have peaks in the 0 -- 0.5 cycles/$P$ range.\\

\noindent
{\bf Positive Drifting (PD)}: In this scenario the phase across the
pulse window changes with a negative slope from the leading to the
trailing edge of the profile, as seen in pulsar J2305$+$3100 (see
figure \ref{fig_phasevar}).  The subpulses show a gradual shift
towards the trailing edge.  The peaks in the HRFS is reflected in case
of PD, with any $f_p$ in LRFS seen at {\it\={f}$_p$}(=1-$f_p$) in the
HRFS, i.e. the HRFS peak lies in the 0.5 -- 1 cycles/$P$ range.\\

\noindent
{\bf Amplitude Modulated Drifting (AMD)}: The AMD corresponds to
pulsars where the subpulses do not move across the pulse window but
exhibit periodic change in intensity.  The phases across the pulse
window do not show much change and mostly remains flat with near zero
slope as seen in pulsar J1919+0021 (figure \ref{fig_phasevar}).  Any
$f_p$ seen in LRFS appears at two locations corresponding to $f_p$ and
{\it\={f}$_p$} in the HRFS with equivalent $S$.

The phase behaviour in pulsar J1034$-$3224 (figure \ref{fig_phasevar})
is complicated without any clear trend.  The leading component of the
profile shows a slightly positive slope which abruptly changes to a
more pronounced negative slope in the later components.  The peaks in
the HRFS is similar to the AMD case with two prominent peaks having
equivalent $S$.  We have not classified this pulsar under any of the
three schemes.

It should be noted that the three cases described above correspond to
different morphologies of the pulsar profile \citep{ran86,ran93a}.
The ND and PD belong to the category of phase modulated drifting and
is mostly seen in pulsars which have conal components and are
typically classified as conal single or double.  The AMD is usually
found in pulsars with core components.  Table \ref{drift_list} lists
the ND and PD as phase modulated drifting and AMD under amplitude
modulation.  In table \ref{drft_par} we have indicated the phase
nature of drifting in each pulsar.

\section{\large Physical Properties} \label{sec:cor}
\begin{figure*}
\begin{tabular}{@{}lr@{}}
{\mbox{\includegraphics[angle=0,scale=0.63]{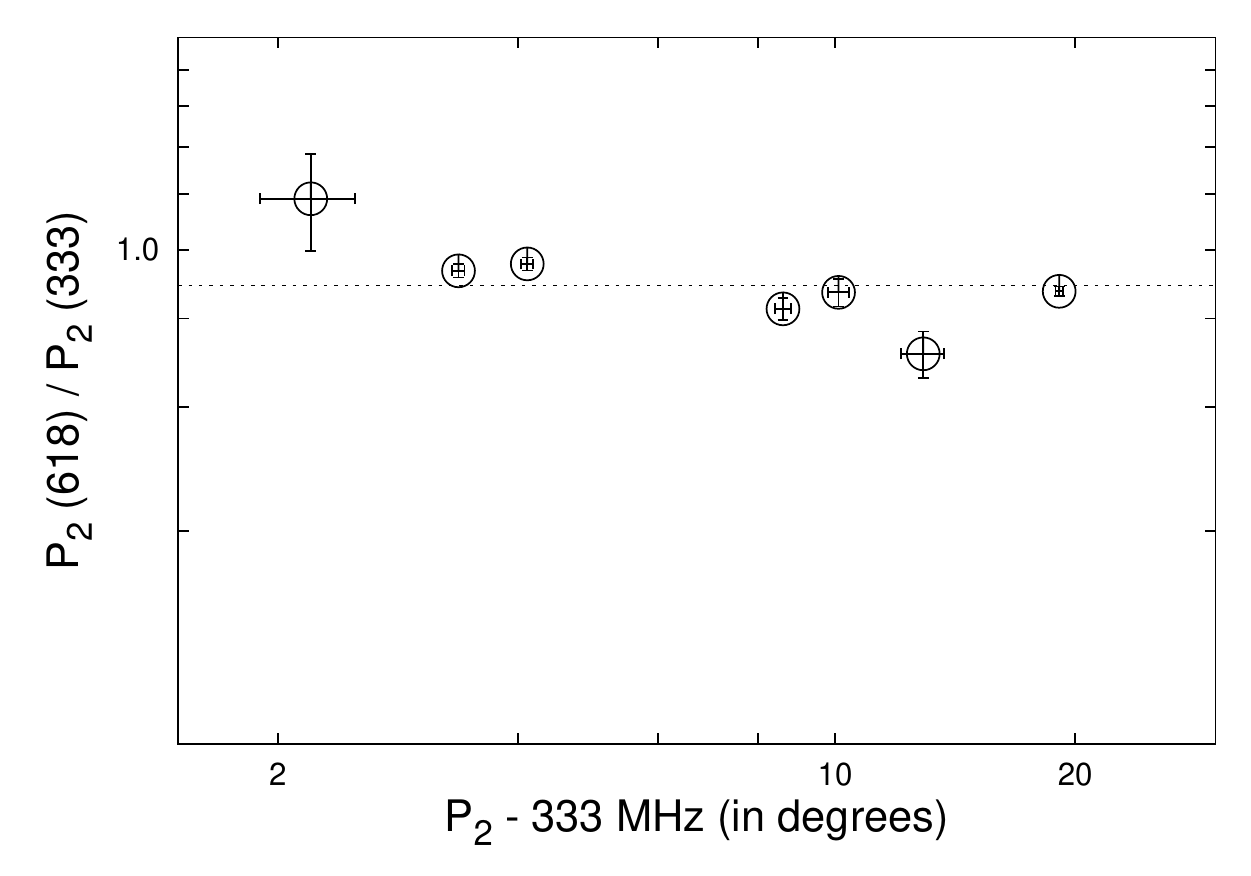}}} &
{\mbox{\includegraphics[angle=0,scale=0.63]{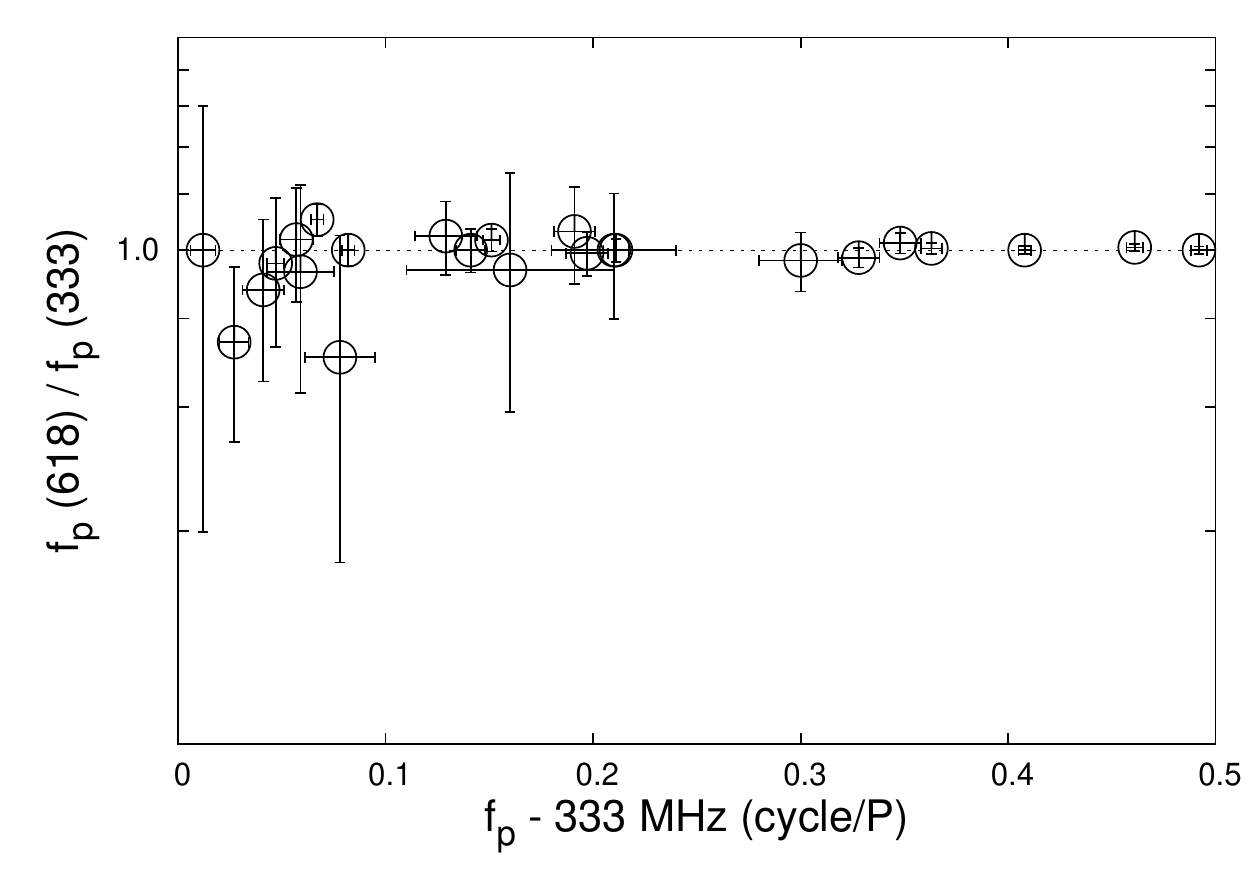}}} \\
\end{tabular}
\caption{The figure shows the frequency dependency of $P_2$ (left
  panel) and $f_p$ (right panel).  In each case we plot the ratio of
  618 and 333 MHz measurement as a function of 333 MHz value.  The
  horizontal (dashed) line in the left panel corresponds to the best
  fit value 0.89 implying that the $P_2$ at 618 MHz is lower than 333
  MHz.  This is expected from radius to frequency mapping where the
  profile width and component separation decreases with increasing
  frequency.  The horizontal (dashed) line on the right panel is set
  at 1 and shows that within errors the measurements at the two
  frequencies are identical.
\label{fig_Pcorr}}
\end{figure*}

\subsection{Frequency Dependence}
\noindent
The separate measurements at two frequencies, 333 MHz and 618 MHz,
allowed us to investigate the frequency dependence of drifting.  There
were 100 pulsars observed at the two frequencies out of which 25
pulsars showed drifting at both frequencies.  12 pulsars exhibited
drifting at 333 MHz and not 618 MHz while 9 pulsars only at 618 MHz.

We were able to measure $P_2$ at both frequencies in 7 pulsars.  In
order to understand the frequency dependence of $P_2$, we determined
the ratio between 618 MHz and 333 MHz and plotted it as a function of
333 MHz value (figure \ref{fig_Pcorr}, left panel).  The 618 MHz
measurements were usually lower than the 333 MHz value with an average
ratio of $\sim$0.89 (dashed horizontal line).  This is consistent with
the radius to frequency mapping which states that the pulse width and
component separation decreases with increasing frequency.  Assuming a
power law dependence of $P_2$ on frequency, $P_2 \propto \nu^a$, the
spectral index was $a \sim-$0.19$\pm$0.04.  The spectral index matched
the frequency dependence of pulse width reported in MSPES.

We also explored the frequency evolution of $f_p$ in the 25 pulsars
where dual frequency measurements were available.  We once again
determined the ratio between 618 MHz and 333 MHz and plotted them as a
function of the 333 MHz value (figure \ref{fig_Pcorr}, right panel).
In the figure the horizontal line is drawn around unity and all
ratios, within errors, fall on this line.  This agrees with previous
studies by \citet{now82,wel07} which claim drifting to be a broadband
phenomenon independent of observing frequency.

\subsection{Distribution with Spin down Energy loss}\label{sec:dist}
\begin{figure*}
\begin{center}
\includegraphics[angle=0,scale=0.8]{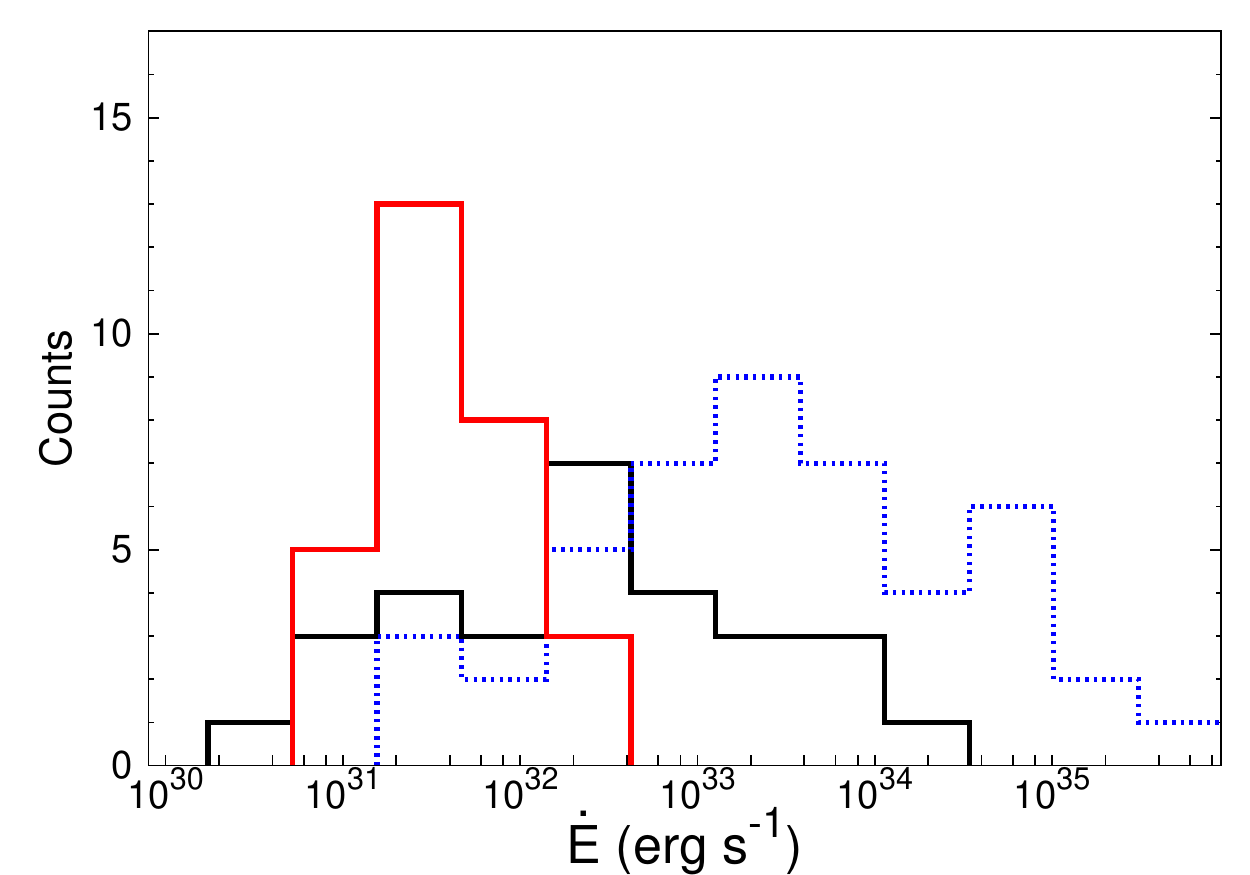}
\caption{The figure shows the distribution of spin-down energy loss
  ($\dot{E}$) for the pulsars used for drifting studies.  The
  histograms are shown for the three distinct populations, the pulsars
  without any periodic features (dotted blue line), the pulsars
  showing phase stationary amplitude modulation (solid black line) and
  the pulsars exhibiting phase modulations (solid red line).  Pulsars
  without any periodic features tend to have higher $\dot{E}$ compared
  to pulsars which show drifting.  The figure also shows that the
  phase modulated drifting is seen only in a narrow range of $\dot{E}
  <$ 10$^{33}$ erg~s$^{-1}$.
\label{fig_edotdist}}
\end{center}
\end{figure*}

\noindent
We estimated the distribution of the three groups of pulsars based on
drifting behavior, phase modulated drifting (ND and PD), amplitude
modulated drifting (AMD) and no drifting, as a function of the spin
down energy loss ($\dot{E}$) as shown in figure \ref{fig_edotdist}.
The phase modulated distribution with 28 pulsars is represented in red
histograms and occupy a very narrow region in the lower $\dot{E}$
range, between 10$^{30}$ - 10$^{33}$ erg~s$^{-1}$, with peak around
10$^{31}$ erg~s$^{-1}$.  The amplitude modulation case with 28
pulsars, shown as black histograms, has a much wider distribution
peaking around 10$^{32}$ erg~s$^{-1}$ and declines beyond
5$\times$10$^{34}$ erg~s$^{-1}$.  The pulsars which do not show
drifting is depicted by the blue dotted histograms.  We have excluded
the pulsars with no detectable single pulses (see MSPES) and the 10
pulsars in table \ref{drift_list} with previously reported drifting
but not detected in our analysis, leaving 46 pulsars in this
distribution.  The distribution peaks around $\dot{E} \sim$ 10$^{33}$
erg~s$^{-1}$ and tends toward the higher $\dot{E}$ range with a lower
cutoff around 10$^{31}$ erg~s$^{-1}$.  It should be noted that our
sample selection criterion in MSPES favours lower $\dot{E}$
($<$10$^{36}$ erg~s$^{-1}$) pulsars which will likely skew this
distribution.

We explored the statistical difference between the three distributions
using the Student's t-test which indicate the probability that two
independent samples of data are from the same population by estimating
the mean of each distribution and calculating a `t' value which is the
weighted difference of the means \citep{pre92}.  To apply the above
method we collected the three distributions taking the logarithm of
the $\dot{E}$ value for every pulsar.  The `t' value was 3.75 for the
phase modulated and amplitude modulated sample with a corresponding
probability of 6.3$\times$10$^{-4}$ for the two to be from the same
distribution.  Here, a probability of 1 corresponds to the two
datasets being identical and 0 corresponds to no overlap.  The
corresponding values between the phase modulated sample and the not
drifting pulsars were `t' value of 10.67 and a likely overlap
probability of 10$^{-6}$.  For the amplitude modulated and not
drifting pulsars the `t' value was 4.12 and the probability of overlap
was 1.2$\times$10$^{-4}$.  The analysis indicate that it is very
likely that the means of the three distributions were very different.
It is remarkable to note that pulsars separate out into three distinct
populations based on drifting behaviour along the $\dot{E}$ axis.

Similar exercises were also carried out with other pulsar parameters
$P$, the spin down rate ($\dot{P}$) and characteristic age ($\tau$),
but no clear difference was seen between the three drifting
populations along any of these parameters.

\section{Drifting Periodicity}
\subsection{Alias Effect}\label{sec:alias}
\noindent
The fluctuation spectral analysis measures the different frequencies
associated with drifting characterised by $f_p$ in the LRFS.  The
different models for the emission mechanism in pulsars are mostly
concerned with the drift periodicity $P_3$, which is the interval
between subpulses to repeat at the same location in the pulse window.
As mentioned earlier, in certain models the subpulse drift
phenomenon is associated with drifting of sparks in the IAR.  The
$P_3$ is then used to estimate the drift velocity of sparks in
the IAR and hence is important for understanding the physical
conditions in the pulsar magnetosphere.  There is however the
aliasing effect that is responsible for uncertainty in estimating the
actual $P_3$ from the measured $f_p$, i.e $P_3$ is not simply 1/$f_p$,
owing to certain limitations of observations.  The pulsed emission is
visible for a short duration every period which makes any measured
frequency subject to certain restrictions.  This ensures that $f_p$, in 
LRFS, at any longitude is only measured in the frequency range 0 - 0.5
cycles/$P$.  The aliasing effect ensures that any periodicity outside
this frequency range will be folded into the above domain.

The aliasing effect associated with $f_p$ is whether the measured
frequency is the actual frequency or the reflection of the larger 
fundamental frequency in the measured window.
This means that for any actual frequency, $f^a_p$ $>$ 0.5 cycles/$P$, the 
measured $f_p$ will be different.  The two frequencies are related as 
\citep{gup04}
\begin{equation}
f^a_p = 2k\times0.5 + (-1)^lf_p 
\end{equation}
Here, $k$ = $INT[(n+1)/2]$ and $l = mod(n,2)$, with n being the alias order, $n\times$0.5 $< f^a_p <$ $(n+1)\times$0.5 cycles/$P$.

We explore one manifestation of aliases arising as a result of subpulse 
motion across the pulse window and the drift direction. The pulsar signal is 
not restricted to a single longitude but spread over a finite window, 
characterised by the profile width.  This implies that for a given choice of 
$f_p$ there is an inherent ambiguity in the direction of subpulse motion, i.e. 
whether the subpulses are moving from the leading to the trailing edge of the 
profile or in the opposite direction. 
To illustrate this we note that the ND correspond to the subpulses in 
subsequent periods appearing at earlier 
longitudes, and the PD is seen when the subpulses appear at later longitudes.  
If the subpulses are intrinsically moving from the trailing to the leading edge
of the profile, ND would correspond to the actual peak being $f_p$ (0 $< f_p <$
0.5 cycles/$P$) while PD would imply the actual peak to be {\it\={f}$_p$} 
(=1-$f_p$, 0.5 $<$ {\it\={f}$_p$} $<$ 1 cycles/$P$).  The opposite is true if 
the subpulses are moving from the leading to the trailing edge, ND would 
correspond to {\it\={f}$_p$} while PD would correspond to $f_p$.  
Unfortunately, any observation of the pulse window gives a snapshot of the 
subpulse structure with no way to estimate their actual direction of motion.
In appendix \ref{sec:simulate_alias} we have simulated single pulses to 
highlight the degeneracy associated with drift direction.
The salient features of the aliasing problem based on the direction of
subpulse motion is summarized in table \ref{tab_alias}.

\begin{table*}
\begin{center}
\caption{Schematic of Aliasing associated with Drift direction.
\label{tab_alias}}
{\begin{tabular}{ccc}
\tableline
 Drifting Type & 1$^{st}$ alias ($P_3 >$ 2$P$) & 2$^{nd}$ alias ($P_3 <$ 2$P$) \\
\tableline
                        &                                         &                                         \\
 Negative Drifting (ND) &  subpulse from trailing to leading edge &  subpulse from leading to trailing edge \\
                        &                                         &                                         \\
 Positive Drifting (PD) &  subpulse from leading to trailing edge &  subpulse from trailing to leading edge \\
                            &                                         &                                         \\
 Amplitude Modulation (AMD) & line of sight at lower part of subpulse & line of sight at upper part of subpulse \\
\tableline
\end{tabular}}
\tablenotetext{~}{\normalsize The pulsar rotation direction is from the leading to the trailing edge.}
\end{center}
\end{table*}

\subsection{Determining $P_3$}
\noindent
As argued above the actual drifting periodicity is intrinsically indeterminate,
especially whether $f_p$ is the actual frequency or a reflection of the 
larger fundamental frequency in the measured window. Hereafter, we assume the 
measured $f_p$ to be the fundamental frequency which physically corresponds to 
the frequency of spark repetition. 
In case of the phase modulated drifting (ND and PD) it is possible to determine the alias between $P$ and 2$P$  if the direction of subpulse motion is known as we discuss below.
As discussed in the introduction the subpulse motion is closely linked 
to the sparks formed in the IAR.  The sparks in the IAR has a specific 
direction, lagging behind the co-rotation of the neutron star.  Every 
subsequent period, as the pulsar comes within our field of view, the spark 
associated plasma are lagging behind and gives an impression of moving opposite
to the co-rotation.  If the above hypothesis is correct we have a preferred 
direction for drifting. The corotation direction is intrinsically from the 
leading to the trailing edge of the profile, which implies that the subpulses 
are expected to move in the opposite direction, i.e from the trailing to the 
leading edge.  As explained in table \ref{tab_alias}, this gives a preferred 
periodicity for the phase modulated drifting, ND with $P_3$ = 1/$f_p$ and PD 
with $P_3$ = 1/(1-$f_p$).  In other words the   drift periodicity is reflected 
in the peaks of the HRFS. We have determined the unaliased $P_3$ values in the phase modulated pulsars
as shown in table \ref{drft_par}, which also lists the unresolved
$P_3$ = 1/$f_p$ for the amplitude modulated cases.  In each $P_3$
measurement we used the weighted average of the $f_p$ measurements
from the two frequencies.  The error is estimated as $\delta$$P_3$ =
$\delta$$f_p$/$f_p^2$.

 The definition of $P_3$ in this work is different from many
  studies in the literature where the drifting even in non-aligned
  pulsars is believed to originate as a result of circulating beamlets
  around the magnetic axis \citep{gil00}.  In our assumption of
  drifting the direction of subpulse motion is independent of whether
  the line of sight is along the inner  
or outer\footnote{The inner line of of sight, often 
called as negative $\beta$ is when 
the observer cuts the emission beam between the rotation axis
and the magnetic axis, and outer line of sight, or positive 
$\beta$ is when the sightline cuts the beam outside the rotation 
and magnetic axis.} region of the emission
  beam.  This is contrary to the assumptions in other works,
  e.g. \citet{des99,wel06,wel07}, where the inner and outer line of
  sights have opposite directions of subpulse motion.  As a result
  most of these works estimate $P_3$ = 1/$f_p$ for both the ND and PD
  \citep{wel06,wel07} cases with the direction of drifting associated
  with the inner or outer line of sight.  To summarize, our estimation
  of $P_3$ is identical to \citet{wel06,wel07} for ND and AMD pulsars.
  In case of PD pulsars our estimations $P_3$ = 1/(1-$f_p$) ($P < P_3
  <$ 2$P$) while \citet{wel06,wel07} uses $P_3$ = 1/$f_p$ ($P_3 >$
  2$P$).

\section{\large Effect of Pulsar Parameters on $P_3$}\label{sec:model}
\subsection{Dependence on $\dot{E}$} \label{sec:P3dotE}
\begin{figure*}
\begin{center}
\includegraphics[angle=0,scale=0.75]{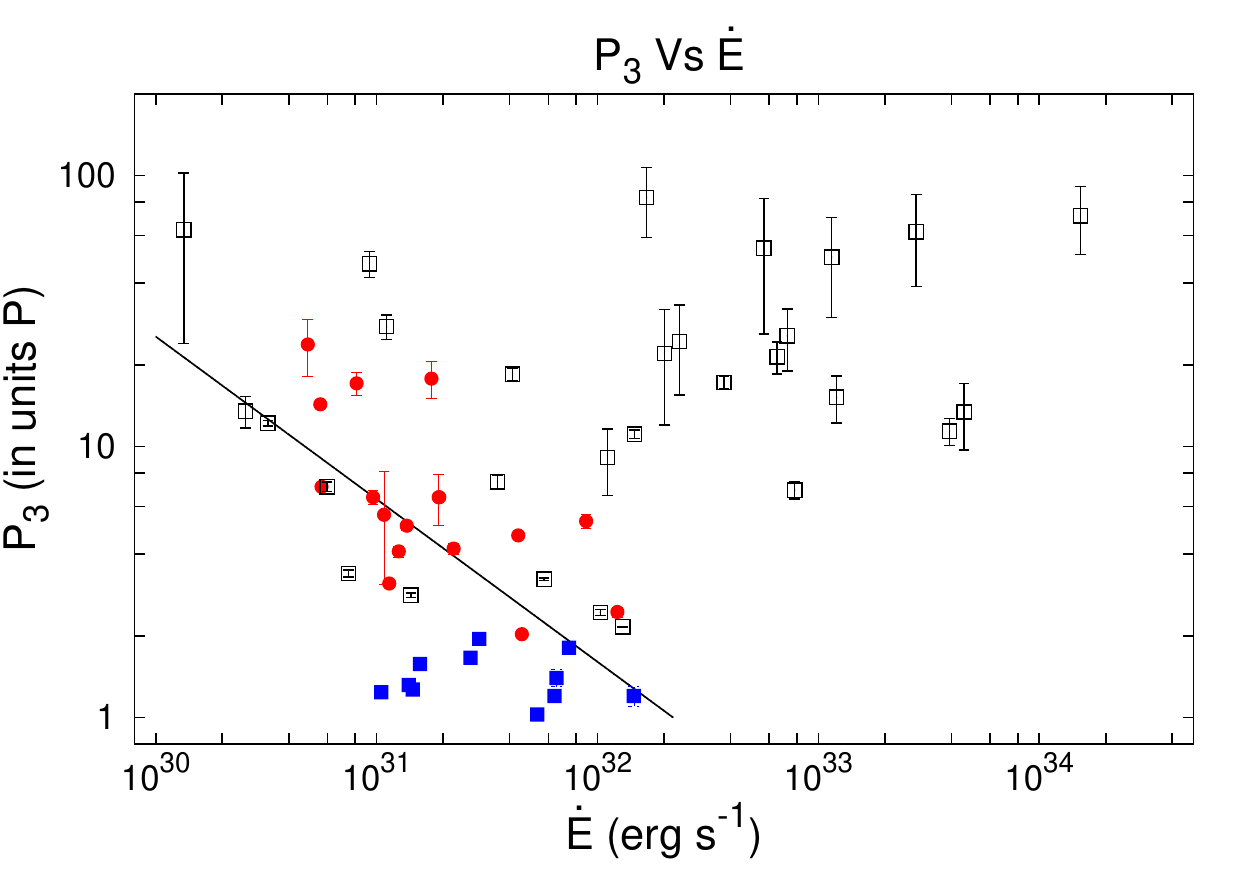}
\caption{The figure shows the variation of the predicted drifting
  periodicity $P_3$ as a function of the spin-down energy loss
  ($\dot{E}$).  The three different populations of periodic features
  are colour coded, with the red dots corresponding to positive phase
  drifting ($P_3 > $2$P$), the blue squares representing the negative
  phase drifting ($P_3 < $2$P$), and the open black squares associated
  with the amplitude modulated drifting ($P_3 > $2$P$, aliasing cannot
  be resolved).  The phase modulated drifting (red and blue points)
  show an anti-correlation with $\dot{E}$.  The phase modulated
  drifting is not seen for $\dot{E} >$ 10$^{33}$ erg~s$^{-1}$, which
  is also the region where the anti-correlation expects $P_3$ to go
  below $P$.
\label{fig_edot_P3}}
\end{center}
\end{figure*}

\noindent
We have determined the variation of $P_3$ (in units of pulsar period
$P$) as a function of $\dot{E}$ ergs/s, as shown in figure
\ref{fig_edot_P3}.  The ND is represented with red dots, the PD with
blue squares and the AMD with black open square symbols.  The phase
modulated drifting show an anti-correlation between $P_3$ and
$\dot{E}$ with a correlation coefficient of --0.7.  The
anti-correlation indicates that the expected $P_3 < P$ for $\dot{E} >$
10$^{33}$ erg~s$^{-1}$.  This is a likely explanation for the absence
of phase modulated drifting at the high $\dot{E}$ regime.  In
  order to determine the relationship between $P_3$ and $\dot{E}$ a
  fit was sought \citep{pre92} with the functional relationship
\begin{equation}
P_3 = \left(\frac{\dot{E}}{\dot{E}_0}\right)^{\delta}
\label{P3_Edot}
\end{equation}
The best fit values corresponded to $\dot{E}_0$ = 
2.3$\pm$0.2 $\times$10$^{32}$~erg~s$^{-1}$ and $\delta$ = --0.6$\pm$0.1 
and is represented by the black line in figure \ref{fig_edot_P3}.

\subsection{Implications for Physical Models}

\noindent
  The key assumption that allowed us to plot
  figure~\ref{fig_edot_P3} is that subpulse drifting results due to
  lack of corotation of sparks, with the velocity of the sparks lagging
  behind the corotation velocity.  In this section we explore
  different IAR models that provide this circumstance.

\subsubsection{The inner vacuum gap model of RS75}
\label{s1}
\noindent
The lack of corotation of sparks was primarily put forward by RS75, where 
they estimated the spark repeating timescale $P_3$, based on two assumptions\\ 
1. they considered an anti-pulsar system, where the pulsar rotation axis and
the magnetic axis are exactly aligned, but are opposite to each other,\\ 
2. and that a number of sparks $n_{\rm sp}$ rotate around the
rotation axis (and hence for an aligned rotator the sparks rotation is
also around the magnetic axis) in a circular path around the
circumference of the polar cap, where the circulation time of each
spark was given as $\hat{P}_3 = n_{sp}\times P_3$. \\ 
This led RS75 to find $P_3 \simeq (5.6/n_{\rm sp})~B_{12}/{P}^2$ (equations 
33 and 34~in RS75), where $B_{12}$ is the magnetic field in units of
10$^{12}$G.  The dependence of $\dot{E}$ on $P_3$ in RS75 model can
be found by using the dipolar magnetic field as $B = 3.2 \times
10^{19} (P\dot{P})^{0.5}$~G with the spin down energy loss, 
$\dot{E} = 4\pi^2~I (\dot{P}/P^3)$, where $I \simeq$ 10$^{45}$ gm~cm$^2$, 
is the moment of inertia of the neutron star. The dependence is given as
\begin{equation}
    P_3 \simeq \frac{5.6}{n_{\rm sp}} \left(\frac{\dot{E}}{\dot{E}_1}\right)^{0.5}.
    \label{P3_RS75}
\end{equation}
Here $P_3$ is expressed in units $P$, $n_{\rm sp}$ corresponds to the
number of sparks in the IVG and $\dot{E}_1 \simeq
$4$\times$10$^{31}$~erg~s$^{-1}$.

Applying the above dependence of $P_3$ with $\dot{E}$ to figure~\ref{fig_edot_P3} 
is difficult since $P_3$ in equation~\ref{P3_RS75} is derived for an aligned rotator which does 
not correspond to a real pulsar. In the non-aligned case if drifting 
is related to sparks that only lag the corotation velocity then the 
direction of the spark velocity only lags behind the observers line 
of sight or in other words the drift motion is around the rotation axis. 
An external observer would essentially see that the sparks are drifting 
across the polar cap, and as a consequence equation~\ref{P3_RS75} which 
is derived by calculating the spark circulation time around the polar 
cap cannot be used to calculate $P_3$. In fact, if the requirement for 
the drifting phenomenon is that the spark has to lag behind corotation, 
then a circular motion around the polar cap for a non-aligned pulsar will
contradict our basic assumption as parts of the spark motion in the
polar cap will lag the corotation velocity while there will be parts
which will be leading.

\subsubsection{Partially Screened Gap Model} \label{s2}
\begin{figure*}
\begin{center}
\includegraphics[angle=0,scale=0.7]{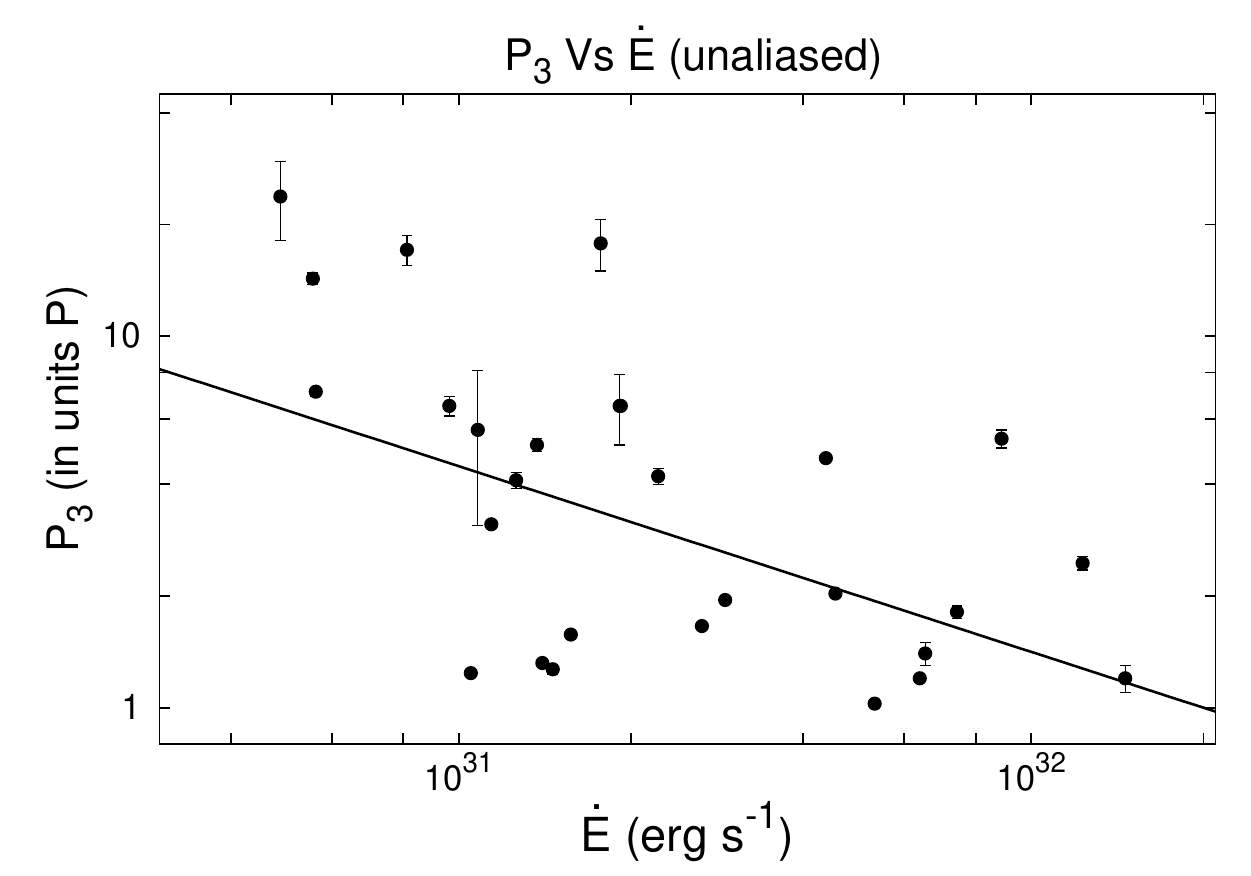}
\caption{The figure shows the estimated $P_3$ of pulsars showing phase
  modulated drifting as a function of $\dot{E}$.  The black line corresponds to 
  equation~\ref{P_3_eta} which gives the behaviour of $P_3$ with $\dot{E}$ 
  as given by the PSG model (see section~\ref{s2} for further details)} .
\label{fig_model_P3}
\end{center}
\end{figure*}
\noindent
We now discuss the IAR model of subpulse drifting for a real pulsar, 
i.e. for a non-aligned case, where 
the spark moves along the locus of the observers line of sight 
at a velocity which is slower than the observers line of sight velocity.
Further, based on several observational constraints
as discussed in the introduction, we will invoke the model where the 
IAR is a PSG. The PSG model 
differs from the IVG by considering a steady flow of
ions from the stellar surface which screens the accelerating electric
field of the gap by a screening factor $\eta$ (where $\eta = 1 -
\rho_i/\rho_{\rm GJ}$, $\rho_i$ is the density of ions in IAR and
$\rho_{\rm GJ}$ is the Goldreich-Julien density; \citealt{gil03a}).
      One simple way of realising the effect of $\eta$ on the
      velocity of sparks is the following. The magnitude of the
      corotational velocity $v_{cr} = (E/B) c$, where $E$ and $B$
      correspond to the electric and magnetic field in the IAR. The
      effect of $\eta$ is to reduce the electric field in the gap such
      that the real drift velocity of the sparks $v_r \approx \eta
      v_{cr}$.
Next we will conjecture that the PSG at any given time is
packed with a number of circular sparks, where the size of a spark and the 
space between the sparks is $h$. This implies that the distance
between the centers of two adjacent sparks is 2h. The sparks move along the 
line of sight of the observer, and hence the spark repeating time at any given 
longitude is estimated as $P_3 = 2 h / v_{d}$.  
It can be shown that if screening factor $\eta$ is small ($\sim$ 0.1) then
$P_3$ is given as:
\begin{equation}
    P_3=\frac{1}{2\pi \cos(\alpha)}\frac{1}{\eta} 
    \label{P_eta}
\end{equation}
Here, $\alpha$ is the angle between the magnetic and rotation axis
(see equation 3.55 Szary 2013).

To find a relation between $P_3$ and $\dot{E}$ in the PSG model,
we first note that the full energy outflow from the polar cap can be
expressed as:
\begin{equation}
    L_{\rm PSG}\simeq\gamma_0 m_{e}c^3\eta n_{\rm GJ}A_{\rm pc}.
    \label{L_PSG}
\end{equation}
Here $\gamma_0$ is a characteristic Lorentz factor of electrons or
positrons accelerated in the gap, $n_{\rm GJ}$ is GJ particle density
and $A_{\rm pc}$ is the area of the polar cap surface.  Let us note
that the value of $n_{\rm GJ}A_{\rm pc}$ does not depend on the surface 
magnetic field configuration due to the magnetic flux conservation law. 
We can estimate $n_{\rm GJ} \simeq$ 7$\times$10$^{10} \cos(\alpha)$ 
($\dot P_{-15}/P$)$^{0.5}$~cm$^{-3}$ and $A_{\rm pc} \simeq$ 3$\times$10$^8 
P^{-1}$~cm$^{2}$ for the dipolar polar cap which gives an invariant value of 
$n_{\rm GJ}A_{\rm pc} \simeq$ 2$\times$10$^{19} \cos(\alpha)$ 
($\dot P_{-15}/P^3$)$^{0.5}$~cm$^{-1}$. 
Here $\dot P_{-15}$ is the spin down rate expressed in units of 
10$^{-15}$ s/s.
The spin down energy loss can be expressed as $\dot E \simeq$~4$\times$10$^{31}$~($\dot P_{-15}/P$) erg~s$^{-1}$ $\simeq$ $\dot{E}_1 \dot P_{-15}/P$.
The quantity $m_{e}c^3 \simeq$~2.5$\times$10$^4$ erg~cm~s$^{-1}$.

Now we can estimate a ratio of $L_{\rm PSG}$ and $\dot E$ as:
\begin{equation}
    \xi=\frac{L_{\rm PSG}}{\dot E}\simeq 1.2\times10^{-8}\eta \gamma_0 \cos(\alpha)\left(\frac{\dot{E}}{\dot{E}_1}\right)^{-0.5}
    \label{xi}
\end{equation}
and, therefore, using Equation (\ref{P_eta})
\begin{equation}
    P_3\simeq 2\times10^{-9}\left(\frac{\gamma_0}{\xi}\right)\left(\frac{\dot{E}}{\dot{E}_1}\right)^{-0.5}.
    \label{P_3_eta}
\end{equation}
The parameter $\gamma_0 \sim 10^6$ and is more or less the same for
most radio pulsars \citep{sza15}.  We can also assume that the major
part of $L_{\rm PSG}$ powers the thermal as well as non-thermal X-ray
emission, and as was shown by \citet{bec09} $\xi\simeq 10^{-3}$.  Thus
we can conclude that the dependence of $P_3$ on $\dot E$ expressed by 
equation (\ref{P_3_eta}) is in agreement with the observed data as shown 
in figure \ref{fig_model_P3} (solid line). It should be noted that there 
is a significant spread in $\xi$ reported in the literature (e.g. \citep{kar12}), 
and this can cause the relation given by equation~\ref{P_3_eta} to vary. 
A detailed study of the effect of PSG model on the drifting phenomenon 
will be carried out in a future work.

\section{\large Summary and Discussion} \label{sec:summary}
\noindent
We have carried out a detailed analysis of drifting in a large sample
of pulsars observed as part of MSPES.  Our fluctuation spectral
analysis were able to detect periodic features in 57 pulsars including
22 pulsars where drifting was seen for the first time.  The drifting
was found to be broadband in nature and showed consistent features of
pulsar emission like radius to frequency mapping.  The pulsar
population can be classified into three groups based on drifting, the
pulsars with phase modulated drifting, the pulsars with amplitude
modulated drifting and finally the pulsars with no drifting.  The
three groups show distinct distributions in terms of their $\dot{E}$
values.  The phase modulated drifting is confined to a narrow range
with low $\dot{E}$ (10$^{30}$ - 10$^{33}$ erg~s$^{-1}$) while the
pulsars with no drifting mostly have high $\dot{E}$ values.  The
amplitude modulated drifting distribution peaks in the intermediate
$\dot{E}$ range.  It seems likely that the pulsar energetics influence
the nature of radio emission.  The phase modulated drifting exhibit
the most ordered emission and is associated with the least energetic
pulsars, while the most energetic pulsars do not show any drifting.  A
study with a larger sample of pulsars would be helpful in further
distinguishing the three populations.  The aliasing effect introduces
ambiguity in associating a periodicity to the measured peak in the
fluctuation spectra.  There is also the added uncertainty connected
with direction of subpulse motion in the pulse window which introduces
an additional two way degeneracy in estimating $P_3$.  However, using
physical arguments about charge motion in the inner acceleration
region we were able to predict the direction of subpulse motion in the
pulsars showing phase modulation.  The estimated $P_3$ in these cases
showed an anti-correlation with $\dot{E}$, which seems to favour the
Partially Screened Gap model of the inner acceleration region in
pulsars.  The anti-correlation also provide a natural explanation for
the absence of phase modulated drifting in high energetic pulsars ($>$
10$^{33}$ erg~s$^{-1}$) as the expected $P_3$ will go below $P$ at
this range.  \\\\

\noindent
{\bf Acknowledgments}: We would like to thank Late Prof. Janusz Gil
for his motivation and encouragement to embark on the subpulse
drifting problem.  We thank the referee Patrick Weltevrede for his
comments especially on the aliasing effect which helped to improve the
paper.  We thank Joanna Rankin, Wojciech Lewandowski and Jarek Kijak for
critical comments on the manuscript.  We would like to thank staff of
Giant Meterwave Radio Telescope and National Center for Radio
Astrophysics for providing valuable support in carrying out this
project.  This work was supported by grants DEC-2012/05/B/ST9/03924
and DEC-2013/09/B/ST9/02177 of the Polish National Science Centre.
This work was financed by the Netherlands Organisation for Scientific
Research (NWO) under project ``CleanMachine" (614.001.301).

\input{table1.tex}

\input{table2.tex}

\appendix
\noindent
We have made the plots and data products from our survey freely available to the user.

\noindent
Several of the data products for each pulsar have been archived in the website:\\
\url{http://mspes.ia.uz.zgora.pl/}\\

\noindent
The bulk download of fluctuation spectra figures for each pulsar is available from:\\
\url{ftp://ftpnkn.ncra.tifr.res.in/dmitra/MSPES/MSPESII}\\

\section{Initial data Processing} \label{sec:data_process}
\subsection{Correcting for baseline variations} 
\noindent
The time series data had a background level which varied due to systematics in the telescope system with the possibility of giving rise to spurious periodicites in the drifting analysis. 
As a first step we intended to remove the background level with systematics in the baseline without affecting any intrinsic periodicity in the pulse sequence. 
The time series was folded using the periodicity of the pulsar and the pulse window ($W_p$) was determined. 
A running mean across the time series was calculated, excluding the pulse window, with a mean value set up for roughly every tenth part of the pulsar period. 
A running polynomial fitting function (4$^{th}$ order) was determined using the running mean data which served as the model for baseline. 
Finally, the time series data was modified by subtracting the baseline level calculated for every data point from the fitted polynomial function. 
This resulted in the data with mean baseline level around zero and mostly free of any spurious periodicity.\\

\subsection{Creating pulse stack} 
\noindent
In order to carry out the various drifting analysis the time series data had to be converted into a pulse stack.
The pulse stack is a two dimensional data form, in contrast with the one dimensional time series, where the horizontal axis corresponds to the pulse number and the vertical axis the pulsar longitude.
Since, the pulsar period is not necessarily an integral multiple of time resolution of observation the generation of pulse stack involved re-sampling the time series data. 
The pulsar period was sampled into an appropriate number of bins ($n_p$) and the time series data was rearranged to fill up the bins for every period. 
The data along any bin corresponded to a particular longitude (the entire period represents a cycle of 360\degr). 
The data could now be identified as $S_{ij}$, where $i$ is the index for the period number and $j$ corresponds to the bin number. 
The pulsar profile was constructed by averaging over the period along the bins, i.e. 
\begin{equation}
p_j = \frac{1}{N}\sum_{i=1}^{N} S_{ij},
\end{equation}
where $N$ is number of pulses.
The pulse window ($W_p$), bound by the start and end bins, was determined as the region above the 5$\sigma$ level from the baseline in the average profile ($p_j$).\\

\subsection{Removing RFI and normalization} 
\noindent
The final step in this round of data processing was to identify and edit out data affected by RFI. 
An off-pulse region was identified outside $W_p$ and the mean ($\mu_i$) and rms ($\sigma_i$) was calculated in the the off-pulse region for each pulse. 
The statistics of the $\sigma_i$ was determined across all the pulses and outliers were identified. 
For the subsequent drifting analysis the time ordering of the pulses needed to be preserved so the outlier pulses (with high value of $\sigma_i$ and as a consequence RFI affected) were edited out by replacing all the entries in their bins with zeros. 
The pulsar profile $p_j$ was once again determined after excluding all the pulses affected by RFI and the peak value of the profile $p_j^{max}$ was determined. 
All the data points $S_{ij}$ were divided by $p_j^{max}$ to normalize the average peak value to unity.

\section{Aliasing in Subpulse Drifting}\label{sec:simulate_alias}
\begin{figure*}
\begin{tabular}{ccc}
{\mbox{\includegraphics[angle=0,scale=0.28]{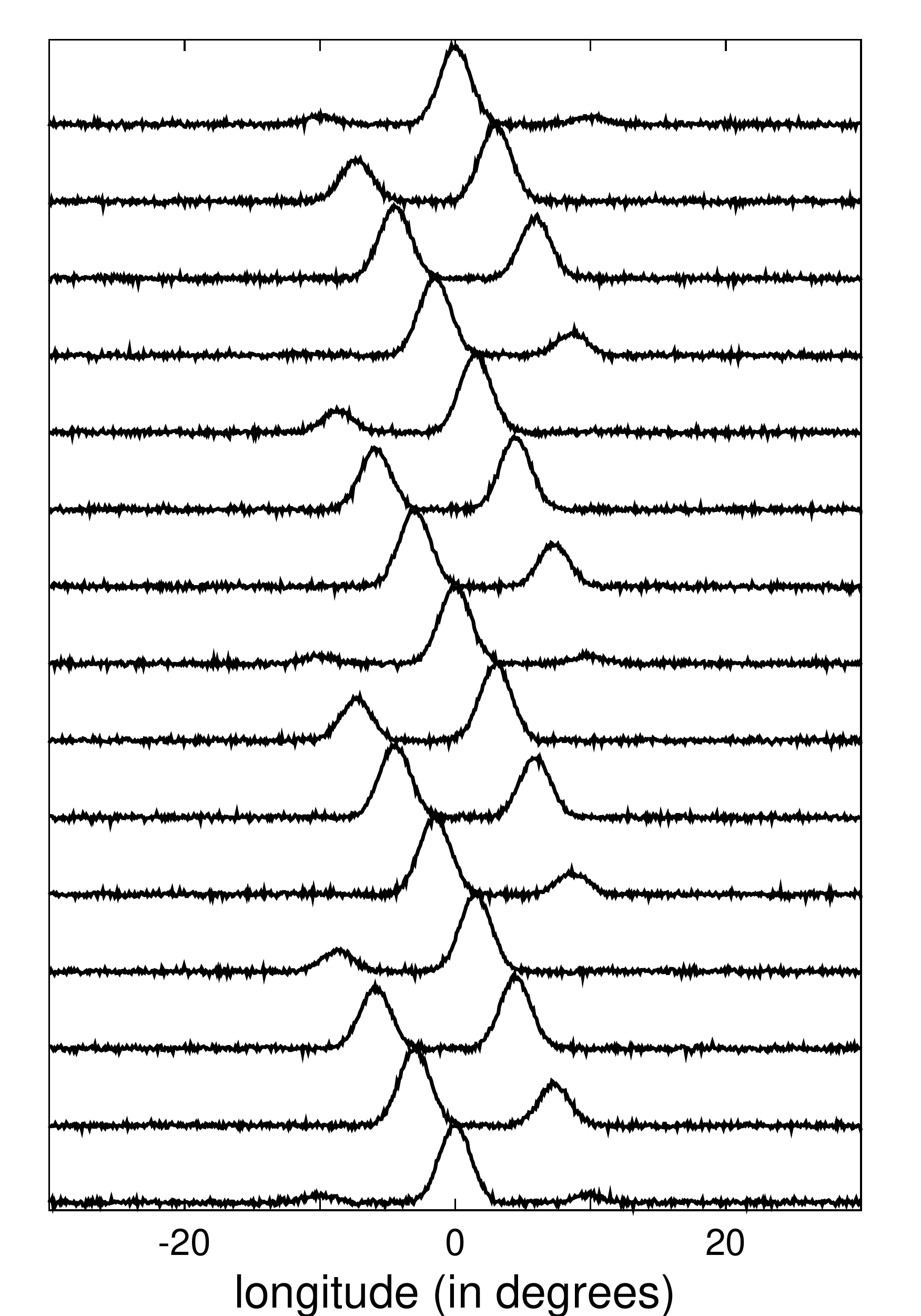}}} &
{\mbox{\includegraphics[angle=0,scale=0.28]{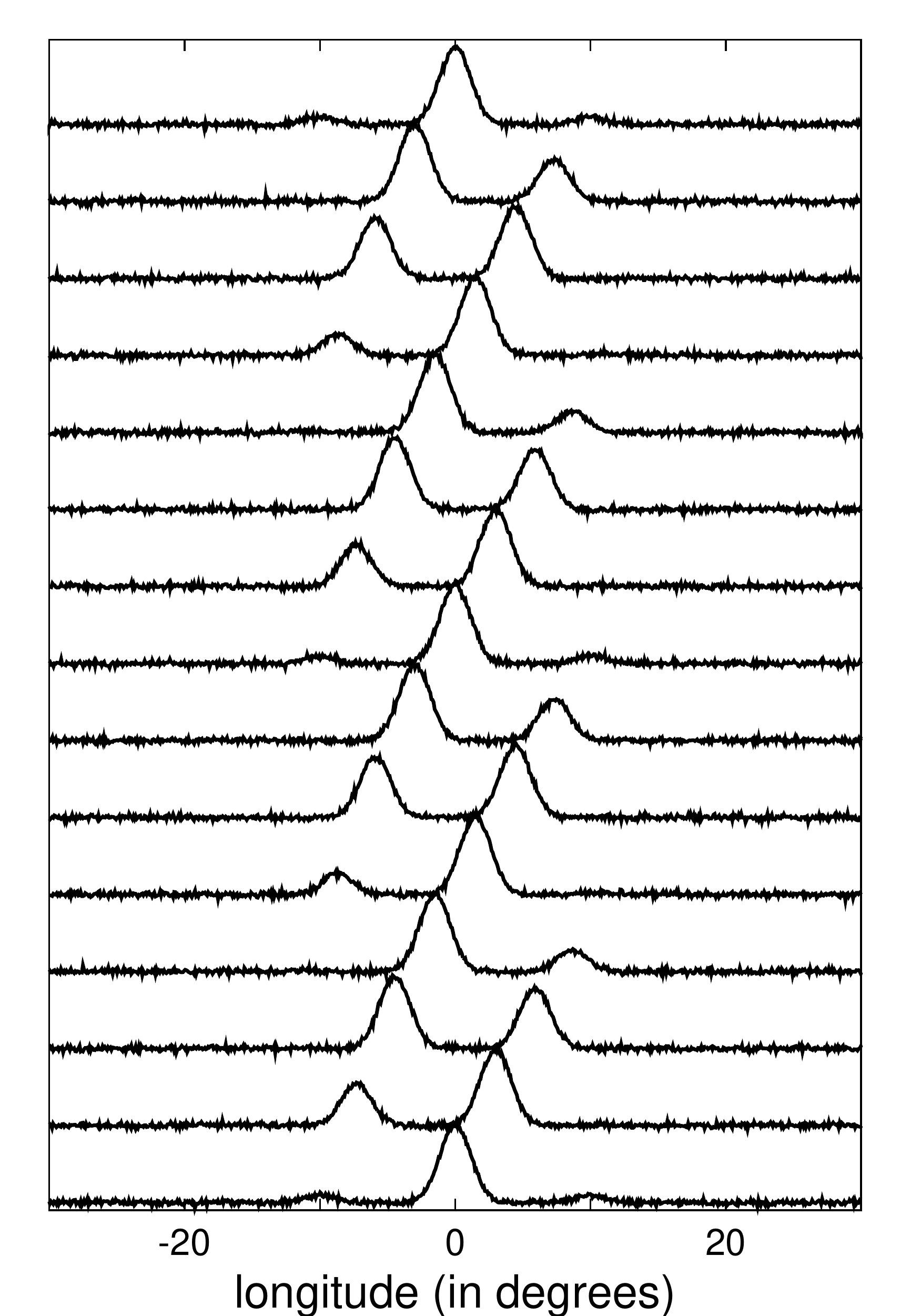}}} &
{\mbox{\includegraphics[angle=0,scale=0.28]{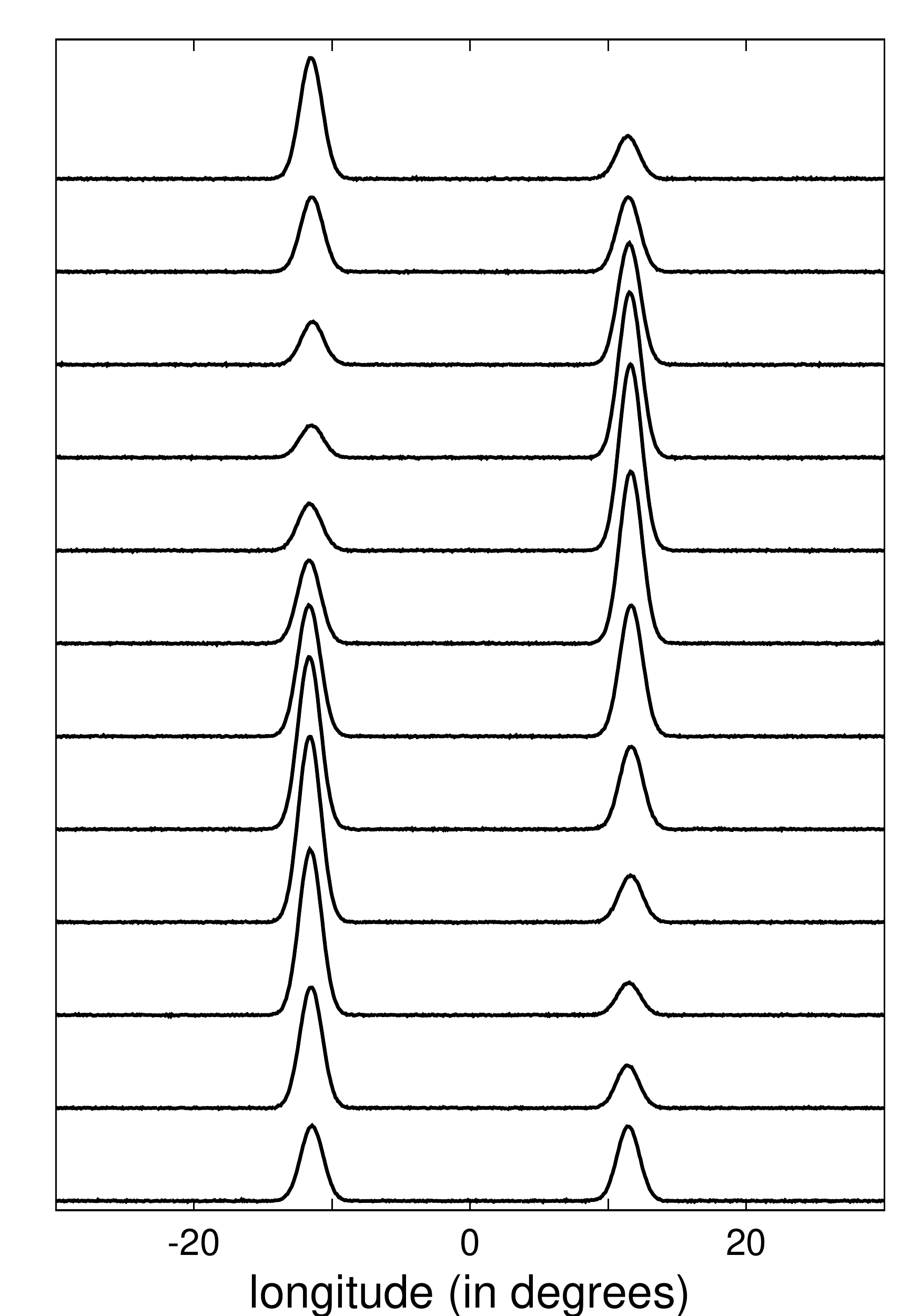}}} \\
\end{tabular}
\caption{The series of simulated single pulses show the three distinct kinds of drifting phenomenon phenomenon seen in pulsars. 
The left panel shows successive pulses appearing at slightly earlier longitudes within the pulse window and corresponds to phase modulated drifting with positive slope (ND).
The central panel shows the successive pulses appearing at slightly later longitudes within the pulse window and corresponds to the phase modulated drifting with negative slope (PD).
The right panel corresponds to the case of amplitude modulation (AMD) where the subpulses do not show any motion across the pulse window but periodic changes in amplitude.
\label{fig_modlsingl}}
\end{figure*}

\noindent
In order to demonstrate the aliasing effect in pulsars we simulated a series of single pulses using the model of \citep{gil03b} to reproduce the drifting phenomenon.
Additionally we have also added gaussian random noise to the simulated single pulses to resemble observations.
The model enabled us to select the drift periodicity $P_3$ as well as the direction of subpulse motion across the pulse window.
We reproduced the three unique drift categories, the negative drifting (ND, figure \ref{fig_modlsingl}, left panel), the positive drifting (PD, figure \ref{fig_modlsingl}, central panel) and the amplitude modulated drifting (AMD, figure \ref{fig_modlsingl}, right panel).
The AMD was simulated as a special geometric configuration where the line of sight traversed the centre of the emission beam.
In each case we carried out the fluctuation spectral analysis on the simulated data.
We explain the result of our analysis for each of these configurations.\\

\begin{figure*}
\begin{center}
\includegraphics[angle=0,scale=0.67]{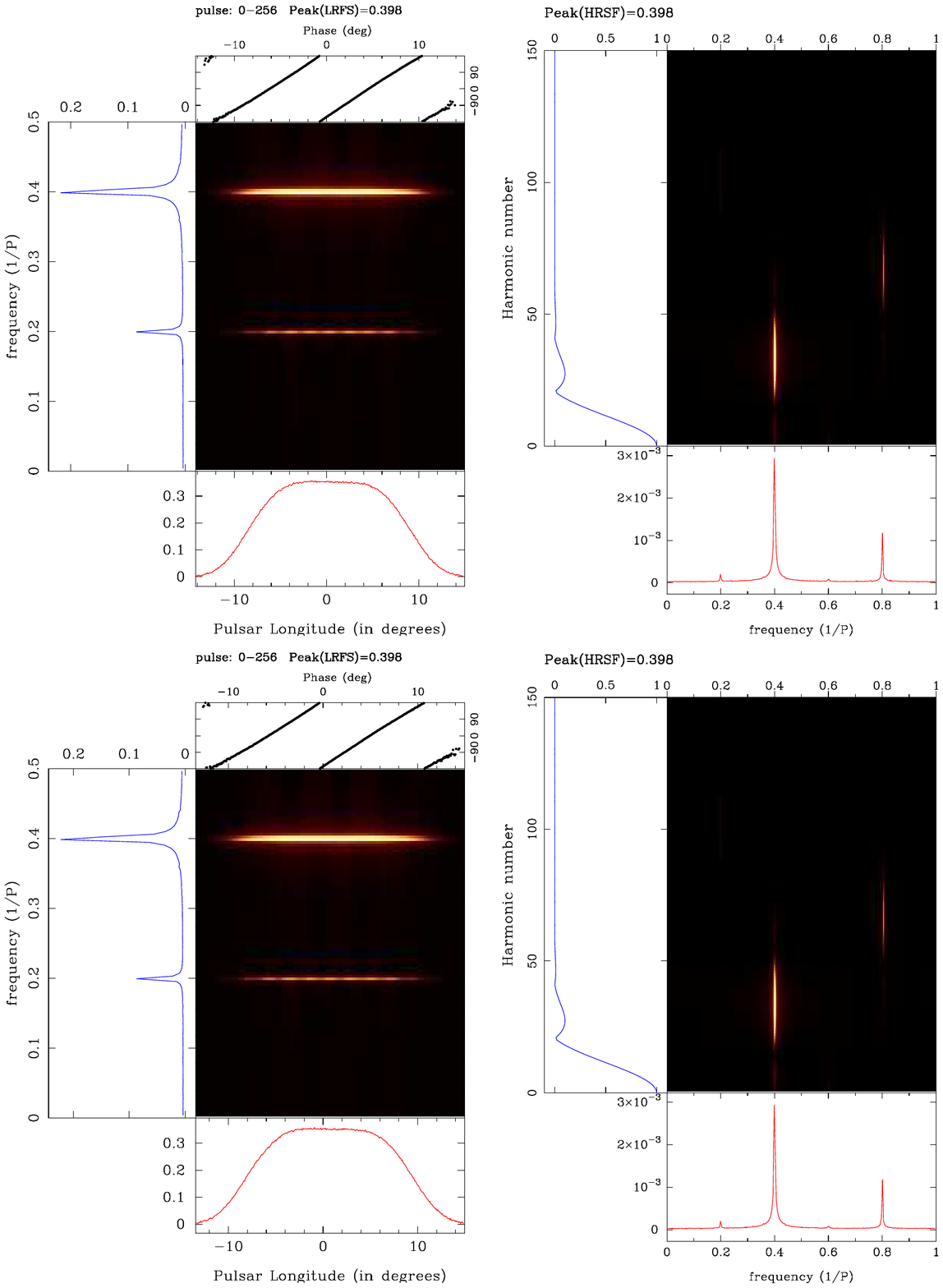}
\caption{The figure show the fluctuation spectral analysis for two simulated data of phase modulated subpulse drifting with positive slope across the pulse window.
The upper panel corresponds to the LRFS (top left) and HRFS (top right) for a drifting periodicity of $P_3$ = 2.5 $P$ and the subpulses moving from the trailing to the leading edge of the profile.
The plots on the lower panel is the corresponding LRFS (bottom left) and HRFS (bottom right) for the drifting periodicity $P_3$ = 1.67 $P$ and the subpulses moving from the leading to the trailing edge of the profile.
The measured peaks from fluctuation spectra is $f_p$ = 0.398 cycle/$P$ in both the data sets including identical second and third harmonics.} 
\label{fig_fs_pos}
\end{center}
\end{figure*}

\noindent
{\bf Phase modulation, positive slope :} As explained in the main text the aliasing effect makes it impossible to distinguish between two periodicities $P_3$ and {\it\={P}$_3$} from the measured peak in the fluctuation spectra, $f_p$, where the periodicities are $P_3$ = 1/$f_p$ and {\it\={P}$_3$} = 1/(1-$f_p$), since there is no way to distinguish whether the subpulses are moving from trailing to leading edge of the profile or vice versa.
In order to demonstrate this we simulated two separate sets of single pulses with the first set having $P_3$ = 2.5 $P$ and the subpulses moving from the trailing to the leading edge. 
The second case have periodicity $P_3$ = 1.67 $P$, with the subpulses moving from the leading to the trailing edge.
In both these cases the expected $f_p$ = 0.4 cycle/$P$ in LRFS as well as HRFS.
The fluctuation spectra was determined for the two datasets and shown in figure \ref{fig_fs_pos}.
The measured peak in the fluctuation spectra is $f_p$ = 0.398 cycle/$P$.
In addition the second and third order harmonics are seen in the LRFS and HRFS in both cases, implying that for these two different drifting behaviour the fluctuation spectra is indistinguishable.\\

\begin{figure*}
\begin{center}
\includegraphics[angle=0,scale=0.67]{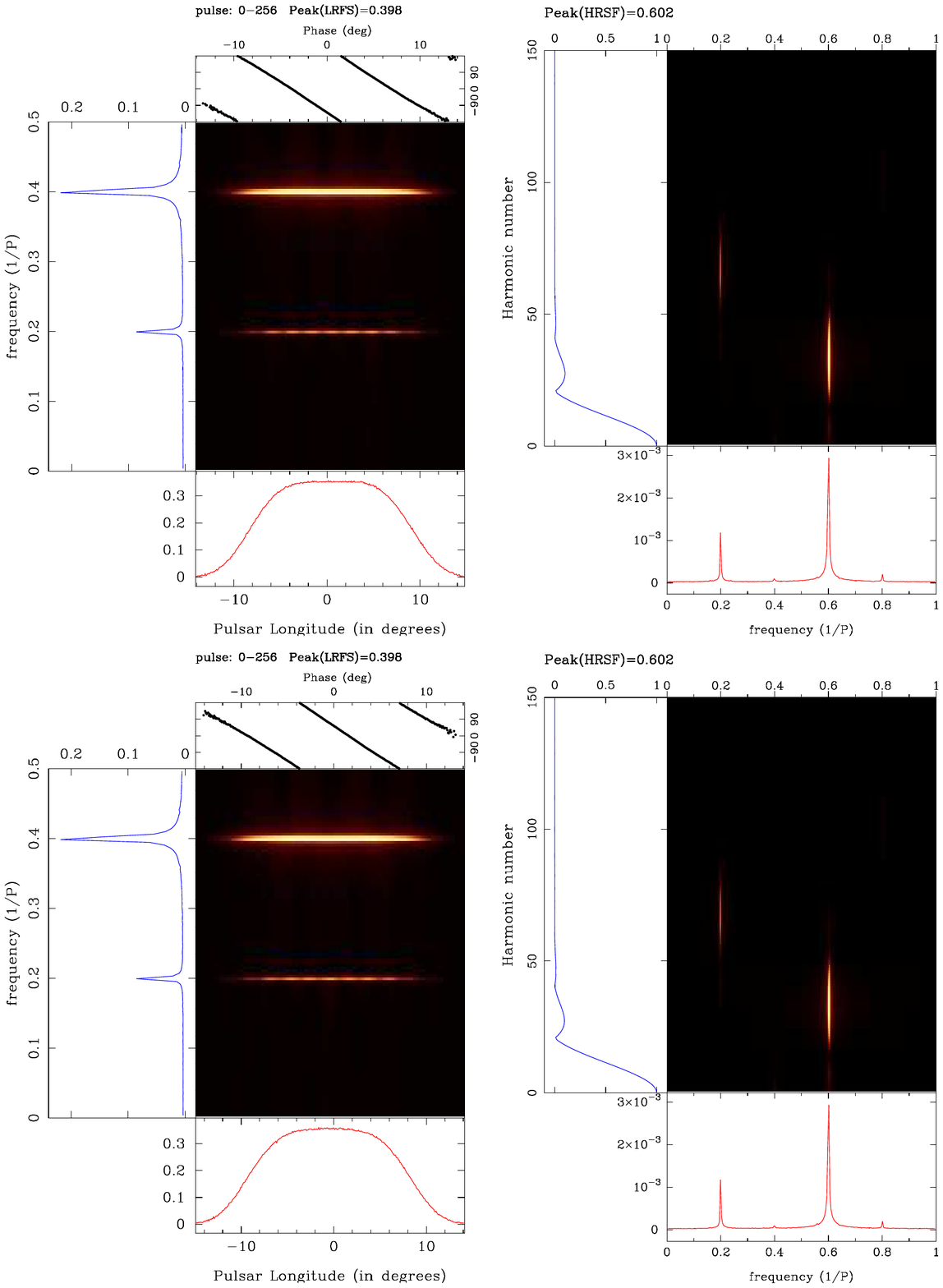}
\caption{The figure show the fluctuation spectral analysis for two simulated data of phase modulated subpulse drifting with negative slope across the pulse window.
The upper panel corresponds to the LRFS (top left) and HRFS (top right) for a drifting periodicity of $P_3$ = 2.5 $P$ and the subpulses moving from the leading to the trailing edge of the profile.
The plots on the lower panel is the corresponding LRFS (bottom left) and HRFS (bottom right) for the drifting periodicity $P_3$ = 1.67 $P$ and the subpulses moving from the trailing to the leading edge of the profile.
The measured peaks from fluctuation spectra is $f_p$ = 0.398 cycle/$P$ for the LRFS and $f_p$ = 0.602 cycle/$P$ in the HRFS for both the data sets including identical second and third harmonics.}
\label{fig_fs_neg}
\end{center} 
\end{figure*}

\noindent
{\bf Phase modulation, negative slope :} We carried out simulations to demonstrate that the fluctuation spectral analysis is identical for drifting periodicities $P_3$ and {\it\={P}$_3$} given the subpulses are moving in opposite directions across the pulse window in the two cases.
Two separate sets of single pulses were generated with the first set having $P_3$ = 2.5 $P$ and the subpulses moving from the leading to the trailing edge.
The second case have periodicity $P_3$ = 1.67 $P$, with the subpulses moving from the trailing to the leading edge.
The expected $f_p$ = 0.4 cycle/$P$ in the LRFS and $f_p$ = 0.6 cycle/$P$ in the HRFS (the negative slope is expected to shift the peak in the 0.5-1 cycle/$P$ range in HRFS).
The fluctuation spectra was determined for the two datasets and shown in figure \ref{fig_fs_neg}.
The measured peak in the fluctuation spectra is $f_p$ = 0.398 cycle/$P$ for the LRFS and $f_p$ = 0.602 cycle/$P$ for the HRFS in both configurations.
In addition the second and third order harmonics are seen in the LRFS and HRFS in both cases, implying that for these two different drifting behaviour the fluctuation spectra is indistinguishable.\\

\begin{figure*}
\begin{center}
\includegraphics[angle=0,scale=0.67]{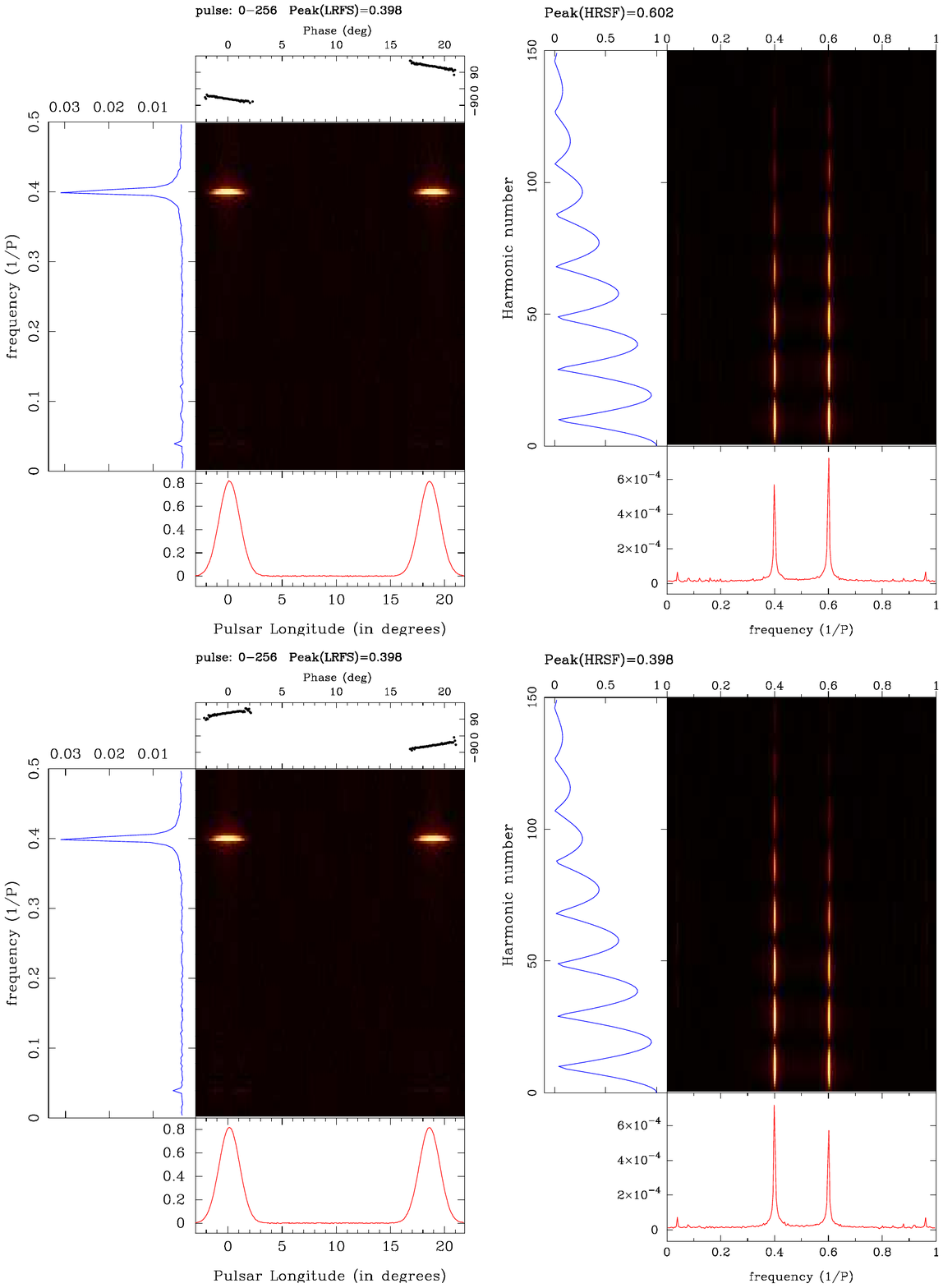}
\caption{The figure show the fluctuation spectral analysis for two simulated data of amplitude modulated subpulse drifting with no subpulse motion across the pulse window. 
The subpulses traverse orthogonal to the line of sight with steady drift periodicities.
The upper panel corresponds to the LRFS (top left) and HRFS (top right) for a drifting periodicity of $P_3$ = 2.5 $P$ while the plots on the lower panel is the corresponding LRFS (bottom left) and HRFS (bottom right) for the drifting periodicity $P_3$ = 1.67 $P$.
The measured peaks from fluctuation spectra is $f_p$ = 0.398 cycle/$P$ in the LRFS, while the HRFS show two strong peaks at $f_p$ = 0.398 cycle/$P$ and {\it\={f}$_p$} = 0.602 cycle/$P$.
\label{fig_fs_amp}}
\end{center} 
\end{figure*}

\noindent
{\bf Amplitude modulation :} We explore the implication of the subpulses moving with drift periodicities $P_3$ and {\it\={P}$_3$} on the fluctuation spectra.
This corresponds to a specific geometric realization where the line of sight traverses the centre of the emission beam.
We have assumed the subpulses to be symmetric in structure.
In this scenario the observed intensity changes as the subpulses move across the line of sight with maximum intensity when the line of sight cuts the central part of the subpulse and the intensity diminishes towards the edges.
The aliasing uncertainty once again comes into play as the $P_3 > $2$P$ catches the subpulse as it is leaving the line of sight which is identical to the $P_3 < $2$P$ case when the subpulse is entering the line of sight.
We have simulated a double peaked pulse profile with the subpulses moving from above to below in the left component and the opposite sense in the right component.
We generated two datasets of single pulses with the first set having $P_3$ = 2.5 $P$ and the second with periodicity $P_3$ = 1.67 $P$.
The fluctuation spectra was determined for the two datasets and shown in figure \ref{fig_fs_amp}.
The measured peak in the fluctuation spectra is $f_p$ = 0.398 cycle/$P$ in the LRFS in both cases.
In the HRFS there were two strong peaks at $f_p$ = 0.398 cycle/$P$ and {\it\={f}$_p$} = 0.602 cycle/$P$.
Depending on the configuration one of the peaks in the HRFS was slightly stronger than the other, but in actual observations with much more noisy data these height differences will not be measurable.
One additional level of degeneracy in the amplitude modulation case is that the results are independent of the direction of subpulse motion, i.e whether the subpulses cross the line of sight from below to above or vice versa.
This is highlighted by the two separate components which have opposite sense of subpulse motion but still show the same frequency peak in the fluctuation spectra.

\end{document}

%% file: table1.tex
\clearpage
\begin{deluxetable}{ccccccccccc}
\rotate
\tabletypesize{\scriptsize}
\tablecolumns{11}
\tablewidth{0pc}
\tablecaption{The table shows the details of the measurement of drifting features in pulsars. The measurements include the peak frequency $f_p$ from the Longitude Resolved Fluctuation Spectra and the Harmonic Resolved Fluctuation Spectra, the strength of peak feature quantified by the parameter $S$ for each peak and the separation between the subpulses specified by $P_2$. Column 2-6 gives the measurements at 333 MHz while columns 7-11 gives the corresponding values at 618 MHz.}
\tablehead{
\colhead{} & \multicolumn{5}{c}{\underline{333 MHz}} & \multicolumn{5}{c}{\underline{618 MHz}} \\
\hline
\multicolumn{1}{c}{PSR} & \multicolumn{1}{c}{$f_p$(LRFS)} & \multicolumn{1}{c}{$S$(LRFS)} & \multicolumn{1}{c}{$P_2$} & \multicolumn{1}{c}{$f_p$(HRFS)} & \multicolumn{1}{c}{$S$(HRFS)} & \multicolumn{1}{c}{$f_p$(LRFS)} & \multicolumn{1}{c}{$S$(LRFS)} & \multicolumn{1}{c}{$P_2$} & \multicolumn{1}{c}{$f_p$(HRSF)} & \multicolumn{1}{c}{$S$(HRFS)} \\
\hline
\colhead{} & \multicolumn{1}{c}{(cycles/$P$)} & \multicolumn{1}{c}{(P)} & \multicolumn{1}{c}{(\degr)} & \multicolumn{1}{c}{(cycles/$P$)} & \multicolumn{1}{c}{(P)} & \multicolumn{1}{c}{(cycles/$P$)} & \multicolumn{1}{c}{(P)} & \multicolumn{1}{c}{(\degr)} & \multicolumn{1}{c}{(cycles/$P$)} & \multicolumn{1}{c}{(P)}}

\startdata
 J0034$-$0721 & 0.150$\pm$0.011 & 23.1$\pm$3.7 & 19.1$\pm$0.2 & 0.150$\pm$0.014 & 23.3$\pm$2.9 & 0.154$\pm$0.011 & 14.4$\pm$2.1 & 16.7$\pm$0.2 & 0.153$\pm$0.012 & 24.0$\pm$3.3 \\
  & 0.301$\pm$0.018 & 2.0$\pm$0.3 & --- & 0.298$\pm$0.023 & 3.3$\pm$0.4 & --- & --- & --- & --- & --- \\
  &  &  &  &  &  &  &  &  &  &  \\
 J0151$-$0635 & 0.067$\pm$0.004 & 33.8$\pm$2.7 & --- & 0.067$\pm$0.004 & 75.5$\pm$5.4 & 0.072$\pm$0.004 & 56.1$\pm$3.1 & --- & 0.072$\pm$0.004 & 92.6$\pm$5.5 \\
  & --- & --- & --- & 0.932$\pm$0.004 & 25.1$\pm$2.6 & --- & --- & --- & --- & --- \\
  &  &  &  &  &  &  &  &  &  &  \\
 J0152$-$1637 & 0.176$\pm$0.031 & 3.0$\pm$0.5 & --- & 0.169$\pm$0.027 & 8.0$\pm$1.1 & 0.169$\pm$0.030 & 2.6$\pm$0.5 & --- & 0.163$\pm$0.029 & 6.1$\pm$0.9 \\
  &  &  &  &  &  &  &  &  &  &  \\
  J0304+1932  & 0.016$\pm$0.009 & 17.8$\pm$3.4 & --- & 0.020$\pm$0.011 & 17.7$\pm$2.7 & --- & --- & --- & --- & --- \\
  & 0.142$\pm$0.039 & 2.4$\pm$0.2 & --- & 0.158$\pm$0.045 & 4.7$\pm$0.3 & 0.157$\pm$0.039 & 2.7$\pm$0.2 & --- & 0.171$\pm$0.049 & 3.0$\pm$0.3 \\
  & --- & --- & --- & 0.986$\pm$0.046 & 3.3$\pm$0.3 & --- & --- & --- & --- & --- \\
  &  &  &  &  &  &  &  &  &  &  \\
 J0525+1115  & --- & --- & --- & 0.684$\pm$0.052 & 1.9$\pm$0.4 & --- & --- & --- & --- & --- \\
  &  &  &  &  &  &  &  &  &  &  \\
 J0630$-$2834 & --- & --- & --- & --- & --- & 0.134$\pm$0.044 & 1.6$\pm$0.3 & --- & --- & --- \\
  & --- & --- & --- & --- & --- & --- & --- & --- & 0.855$\pm$0.032 & 5.1$\pm$0.7 \\
  &  &  &  &  &  &  &  &  &  &  \\
 J0758$-$1528 & --- & --- & --- & --- & --- & 0.042$\pm$0.018 & 5.9$\pm$0.7 & --- & 0.047$\pm$0.022 & 8.7$\pm$1.0 \\
  & --- & --- & --- & --- & --- & --- & --- & --- & 0.953$\pm$0.015 & 15.2$\pm$1.1 \\
  &  &  &  &  &  &  &  &  &  &  \\
 J0820$-$1350 & 0.211$\pm$0.006 & 61.7$\pm$2.3 & 4.11$\pm$0.07 & 0.211$\pm$0.007 & 59.8$\pm$2.4 & 0.211$\pm$0.005 & 65.3$\pm$1.8 & 3.93$\pm$0.05 & 0.211$\pm$0.005 & 69.6$\pm$2.1 \\
  & --- & --- & --- & --- & --- & --- & --- & --- & 0.788$\pm$0.005 & 7.3$\pm$1.2 \\
  &  &  &  &  &  &  &  &  &  &  \\
 J0837+0610  & 0.462$\pm$0.006 & 31.5$\pm$3.6 & --- & 0.458$\pm$0.004 & 50.8$\pm$6.3 & 0.463$\pm$0.007 & 20.7$\pm$2.0 & --- & 0.464$\pm$0.008 & 19.4$\pm$2.9 \\
 & --- & --- & --- & --- & --- & --- & --- & --- & 0.536$\pm$0.004 & 46.0$\pm$5.1 \\
  &  &  &  &  &  &  &  &  &  &  \\
 J0846$-$3533 & 0.493$\pm$0.006 & 19.4$\pm$0.5 & --- & 0.492$\pm$0.005 & 51.9$\pm$1.2 & 0.491$\pm$0.004 & 17.3$\pm$0.9 & --- & 0.492$\pm$0.005 & 50.8$\pm$1.9 \\
  & --- & --- & --- & 0.507$\pm$0.006 & 28.8$\pm$1.0 & --- & --- & --- & --- & --- \\
  &  &  &  &  &  &  &  &  &  &  \\
 J0944$-$1354 & --- & --- & 1.4$\pm$0.3 & 0.156$\pm$0.008 & 19.7$\pm$3.2 & --- & --- & --- & --- & --- \\
  &  &  &  &  &  &  &  &  &  &  \\
 J0959$-$4809 & --- & --- & --- & --- & --- & --- & --- & --- & 0.179$\pm$0.041 & 4.2$\pm$0.4 \\
  &  &  &  &  &  &  &  &  &  &  \\
 J1034$-$3224 & 0.139$\pm$0.007 & 24.4$\pm$1.6 & --- & 0.141$\pm$0.009 & 31.0$\pm$3.0 & 0.139$\pm$0.012 & 5.1$\pm$0.5 & --- & 0.138$\pm$0.013 & 14.4$\pm$1.7 \\
 & --- & --- & --- & 0.859$\pm$0.007 & 37.9$\pm$2.8 & --- & --- & --- & 0.860$\pm$0.013 & 18.4$\pm$1.0 \\
  &  &  &  &  &  &  &  &  &  &  \\
 J1041$-$1942 & --- & --- & --- & --- & --- & 0.233$\pm$0.024 & 2.5$\pm$0.3 & --- & --- & --- \\
  & --- & --- & --- & --- & --- & --- & --- & --- & 0.768$\pm$0.028 & 7.0$\pm$0.6 \\
  &  &  &  &  &  &  &  &  &  &  \\
 J1116$-$4122 & 0.027$\pm$0.010 & 17.5$\pm$3.2 & 2.9$\pm$0.2 & 0.028$\pm$0.011 & 18.6$\pm$3.3 & --- & --- & --- & --- & --- \\
  & 0.050$\pm$0.010 & 13.3$\pm$2.3 & --- & 0.050$\pm$0.012 & 15.6$\pm$2.6 & --- & --- & --- & --- & --- \\
  & --- & --- & --- & 0.950$\pm$0.017 & 9.4$\pm$1.1 & --- & --- & --- & --- & --- \\
  & --- & --- & --- & 0.973$\pm$0.024 & 8.1$\pm$0.8 & --- & --- & --- & --- & --- \\
  &  &  &  &  &  &  &  &  &  &  \\
  J1239+2453  & 0.361$\pm$0.011 & 18.0$\pm$1.2 & --- & 0.360$\pm$0.015 & 19.0$\pm$1.6 & 0.361$\pm$0.014 & 15.6$\pm$1.0 & --- & 0.361$\pm$0.019 & 14.3$\pm$1.5 \\
  & --- & --- & --- & 0.639$\pm$0.011 & 24.6$\pm$1.8 & --- & --- & --- & 0.639$\pm$0.010 & 30.1$\pm$2.6 \\
  &  &  &  &  &  &  &  &  &  &  \\
 J1328$-$4921 & 0.296$\pm$0.018 & 8.5$\pm$0.7 & --- & 0.296$\pm$0.018 & 16.5$\pm$1.4 & 0.294$\pm$0.017 & 4.8$\pm$0.4 & --- & 0.291$\pm$0.018 & 16.1$\pm$1.1 \\
 & --- & --- & --- & 0.705$\pm$0.019 & 15.6$\pm$1.0 & --- & --- & --- & 0.706$\pm$0.018 & 12.5$\pm$0.6 \\
  &  &  &  &  &  &  &  &  &  &  \\
 J1418$-$3921 & --- & --- &  --- & --- & --- & 0.400$\pm$0.004 & 23.2$\pm$1.8 & --- & --- & --- \\
  & --- & --- &  --- & --- & --- & --- & --- & --- & 0.598$\pm$0.005 & 50.3$\pm$3.4 \\
  &  &  &  &  &  &  &  &  &  &  \\
 J1527$-$3931 & 0.026$\pm$0.004 & 33.5$\pm$6.2 & --- & --- & --- & --- & --- & --- & --- & --- \\
  & --- & --- & --- & 0.975$\pm$0.004 & 59.7$\pm$9.8 & --- & --- & --- & --- & --- \\
  &  &  &  &  &  &  &  &  &  &  \\
 J1555$-$3134 & 0.053$\pm$0.012 & 13.5$\pm$1.9 & --- & 0.053$\pm$0.012 & 24.3$\pm$3.3 & 0.055$\pm$0.010 & 16.7$\pm$2.7 & --- & 0.055$\pm$0.012 & 24.8$\pm$3.6 \\
  & 0.096$\pm$0.008 & 18.3$\pm$1.6 & --- & 0.096$\pm$0.010 & 27.0$\pm$3.0 & 0.098$\pm$0.008 & 21.9$\pm$1.8 & --- & 0.098$\pm$0.008 & 35.6$\pm$3.3 \\
  & --- & --- & --- & 0.902$\pm$0.012 & 7.4$\pm$0.6 & --- & --- & --- & --- & --- \\
  & --- & --- & --- & 0.949$\pm$0.013 & 8.5$\pm$0.5 & --- & --- & --- & 0.947$\pm$0.013 & 8.5$\pm$0.4 \\
  &  &  &  &  &  &  &  &  &  &  \\
 J1603$-$2531 & --- & --- & --- & --- & --- & 0.015$\pm$0.006 & 16.8$\pm$4.6 & --- & 0.014$\pm$0.006 & 24.6$\pm$7.3 \\
  & --- & --- & --- & --- & --- & --- & --- & --- & 0.986$\pm$0.009 & 14.9$\pm$2.1 \\
  &  &  &  &  &  &  &  &  &  &  \\
 J1604$-$4909 & --- & --- & --- & --- & --- & 0.020$\pm$0.011 & 8.4$\pm$1.5 & --- & 0.020$\pm$0.010 & 21.3$\pm$4.3 \\
  & --- & --- & --- & --- & --- & --- & --- & --- & 0.980$\pm$0.016 & 13.2$\pm$1.5 \\
  &  &  &  &  &  &  &  &  &  &  \\
 J1625$-$4048 & --- & --- & --- & --- & --- & 0.017$\pm$0.010 & 3.3$\pm$0.5 & --- & 0.016$\pm$0.010 & 17.3$\pm$2.3 \\
  & --- & --- & --- & --- & --- & --- & --- & --- & 0.983$\pm$0.010 & 22.3$\pm$1.2 \\
  &  &  &  &  &  &  &  &  &  &  \\
 J1645$-$0317 & --- & --- & --- & --- & --- & 0.080$\pm$0.030 & 4.9$\pm$0.6 & --- & 0.080$\pm$0.027 & 7.3$\pm$1.0 \\
 & --- & --- & --- & --- & --- & --- & --- & --- & 0.924$\pm$0.025 & 7.1$\pm$1.0 \\
  &  &  &  &  &  &  &  &  &  &  \\
 J1700$-$3312 & --- & --- & --- & --- & --- & 0.444$\pm$0.020 & 1.3$\pm$0.3 & --- & --- & --- \\
  & --- & --- & --- & --- & --- & --- & --- & --- & 0.553$\pm$0.022 & 6.5$\pm$0.9 \\
  &  &  &  &  &  &  &  &  &  &  \\
 J1703$-$3241 & 0.212$\pm$0.031 & 3.8$\pm$0.3 & --- & 0.215$\pm$0.029 & 5.7$\pm$0.7 & 0.211$\pm$0.022 & 6.5$\pm$0.5 & --- & 0.216$\pm$0.022 & 5.5$\pm$0.7 \\
  & --- & --- & --- & 0.814$\pm$0.055 & 4.0$\pm$0.3 & --- & --- & --- & 0.792$\pm$0.027 & 8.7$\pm$0.5 \\
  &  &  &  &  &  &  &  &  &  &  \\
 J1720$-$2933 & 0.408$\pm$0.003 & 90.5$\pm$1.0 & 10.1$\pm$0.3 & 0.408$\pm$0.004 & 102.0$\pm$1.2 & 0.409$\pm$0.003 & 88.2$\pm$1.0 & 8.8$\pm$0.3 & 0.409$\pm$0.004 & 98.4$\pm$1.2 \\
  & 0.183$\pm$0.004 & 9.5$\pm$0.6 & 7.8$\pm$0.3 & --- & --- & 0.183$\pm$0.003 & 17.8$\pm$0.9 & --- & --- & --- \\
  & --- & --- & --- & 0.592$\pm$0.003 & 11.1$\pm$0.9 & --- & --- & --- & 0.589$\pm$0.004 & 7.5$\pm$0.8 \\
  & --- & --- & --- & 0.817$\pm$0.004 & 19.4$\pm$0.7 & --- & --- & --- & 0.817$\pm$0.004 & 31.3$\pm$0.8 \\
  &  &  &  &  &  &  &  &  &  &  \\
 J1722$-$3207 & --- & --- & --- & --- & --- & 0.044$\pm$0.023 & 2.2$\pm$0.3 & --- & 0.044$\pm$0.016 & 9.7$\pm$1.4 \\
 & --- & --- & --- & --- & --- & --- & --- & --- & 0.958$\pm$0.019 & 7.3$\pm$0.7 \\
  &  &  &  &  &  &  &  &  &  &  \\
 J1733$-$2228 & 0.038$\pm$0.025 & 4.1$\pm$0.3 & --- & 0.039$\pm$0.025 & 12.5$\pm$0.8 & 0.048$\pm$0.029 & 0.9$\pm$0.1 & --- & 0.051$\pm$0.026 & 8.0$\pm$0.7 \\
 & --- & --- & --- & 0.973$\pm$0.017 & 12.4$\pm$0.4 & --- & --- & --- & --- & --- \\
  &  &  &  &  &  &  &  &  &  &  \\
 J1733$-$3716 & --- & --- & --- & --- & --- & 0.013$\pm$0.004 & 5.9$\pm$2.2 & --- & 0.015$\pm$0.004 & 25.6$\pm$8.4 \\
  & --- & --- & --- & --- & --- & --- & --- & --- & 0.988$\pm$0.004 & 32.1$\pm$3.3 \\
  &  &  &  &  &  &  &  &  &  &  \\
 J1735$-$0724 & 0.053$\pm$0.019 & 10.5$\pm$1.2 & --- & 0.052$\pm$0.024 & 8.2$\pm$1.0 & 0.046$\pm$0.023 & 3.3$\pm$0.4 & --- & 0.051$\pm$0.016 & 8.0$\pm$1.3 \\
  & --- & --- & --- & 0.949$\pm$0.017 & 14.6$\pm$1.1 & --- & --- & --- & 0.953$\pm$0.025 & 7.3$\pm$0.6 \\
  &  &  &  &  &  &  &  &  &  &  \\
  J1740+1311  & --- & --- & --- & --- & --- & 0.113$\pm$0.029 & 3.6$\pm$0.3 & --- & 0.113$\pm$0.029 & 6.7$\pm$0.6 \\
  & --- & --- & --- & --- & --- & --- & --- & --- & 0.887$\pm$0.028 & 7.8$\pm$0.5 \\
  &  &  &  &  &  &  &  &  &  &  \\
 J1741$-$3927 & --- & --- & --- & --- & --- & 0.105$\pm$0.052 & 2.0$\pm$0.2 & --- & 0.106$\pm$0.043 & 4.2$\pm$0.3 \\
  & 0.029$\pm$0.017 & 3.6$\pm$0.4 & --- & 0.030$\pm$0.018 & 13.1$\pm$1.5 & 0.015$\pm$0.010 & 13.0$\pm$2.5 & --- & 0.015$\pm$0.009 & 26.3$\pm$5.2 \\
  & --- & --- & --- & --- & --- & --- & --- & --- & 0.879$\pm$0.046 & 3.7$\pm$0.3 \\
  & --- & --- & --- & 0.974$\pm$0.018 & 10.5$\pm$0.9 & --- & --- & --- & 0.988$\pm$0.011 & 16.3$\pm$1.2 \\
  &  &  &  &  &  &  &  &  &  &  \\
 J1741$-$0840 & 0.211$\pm$0.029 & 3.3$\pm$0.3 & --- & --- & --- & 0.206$\pm$0.026 & 3.2$\pm$0.3 & --- & --- & --- \\
  & --- & --- & --- & 0.779$\pm$0.023 & 4.8$\pm$0.4 & --- & --- & --- & 0.789$\pm$0.039 & 6.8$\pm$0.5 \\
  & 0.033$\pm$0.006 & 17.5$\pm$3.5 & --- & 0.036$\pm$0.002 & 20.4$\pm$4.1 & 0.033$\pm$0.009 & 9.9$\pm$1.8 & --- & 0.033$\pm$0.009 & 23.4$\pm$4.7 \\
  & --- & --- & --- & 0.965$\pm$0.001 & 11.9$\pm$9.4 & --- & --- & --- & 0.967$\pm$0.009 & 21.2$\pm$1.5 \\
  &  &  &  &  &  &  &  &  &  &  \\
 J1748$-$1300 & 0.148$\pm$0.012 & 4.7$\pm$0.6 & --- & 0.153$\pm$0.017 & 9.7$\pm$1.4 & --- & --- & --- & --- & --- \\
 & --- & --- & --- & 0.851$\pm$0.015 & 10.2$\pm$1.2 & --- & --- & --- & --- & --- \\
  &  &  &  &  &  &  &  &  &  &  \\
 J1801$-$0357 & 0.012$\pm$0.006 & 28.4$\pm$7.8 & --- & 0.012$\pm$0.006 & 40.4$\pm$11.0 & 0.013$\pm$0.006 & 30.5$\pm$9.2 & --- & 0.012$\pm$0.008 & 31.2$\pm$6.2 \\
  & --- & --- & --- & 0.988$\pm$0.006 & 38.9$\pm$2.4 & --- & --- & --- & 0.987$\pm$0.006 & 39.3$\pm$2.3 \\
  &  &  &  &  &  &  &  &  &  &  \\
 J1801$-$2920 & --- & --- & --- & --- & --- & 0.404$\pm$0.013 & 4.1$\pm$0.7 & --- & 0.404$\pm$0.014 & 16.6$\pm$1.6 \\
  & --- & --- & --- & --- & --- & --- & --- & --- & 0.498$\pm$0.008 & 26.0$\pm$2.8 \\
  & --- & --- & --- & --- & --- & --- & --- & --- & 0.593$\pm$0.008 & 26.6$\pm$2.6 \\
  &  &  &  &  &  &  &  &  &  &  \\
 J1816$-$2650 & 0.245$\pm$0.011 & 1.8$\pm$0.3 & --- & 0.240$\pm$0.025 & 9.3$\pm$0.9 & --- & --- & --- & --- & --- \\
  &  &  &  &  &  &  &  &  &  &  \\
 J1822$-$2256 & 0.059$\pm$0.008 & 18.1$\pm$2.5 & 8.6$\pm$0.2 & 0.060$\pm$0.009 & 35.6$\pm$5.6 & 0.057$\pm$0.007 & 26.4$\pm$3.3 & 7.1$\pm$0.2 & 0.057$\pm$0.008 & 39.4$\pm$5.9 \\
  &  &  &  &  &  &  &  &  &  &  \\
 J1842$-$0359 & 0.082$\pm$0.004 & 4.7$\pm$0.8 & --- & 0.082$\pm$0.004 & 63.1$\pm$4.6 & 0.080$\pm$0.003 & 54.1$\pm$2.0 & --- & 0.080$\pm$0.003 & 106.4$\pm$4.7 \\
  & --- & --- & --- & --- & --- & 0.058$\pm$0.004 & 11.8$\pm$1.0 & --- & 0.059$\pm$0.006 & 20.2$\pm$2.3 \\
  & --- & --- & --- & --- & --- & 0.036$\pm$0.006 & 5.7$\pm$1.0 & --- & 0.035$\pm$0.008 & 11.0$\pm$2.5 \\
  & --- & --- & --- & --- & --- & 0.161$\pm$0.007 & 3.0$\pm$0.4 & --- & 0.159$\pm$0.010 & 6.8$\pm$0.8 \\
  & --- & --- & --- & 0.919$\pm$0.004 & 38.6$\pm$3.6 & --- & --- & --- & 0.920$\pm$0.003 & 107.6$\pm$2.6 \\
  & --- & --- & --- & --- & --- & --- & --- & --- & 0.943$\pm$0.005 & 20.8$\pm$1.4 \\
  & --- & --- & --- & --- & --- & --- & --- & --- & 0.84$\pm$0.008 & 8.4$\pm$0.8 \\
  &  &  &  &  &  &  &  &  &  &  \\
 J1848$-$0123 & --- & --- & --- & --- & --- & 0.051$\pm$0.018 & 4.3$\pm$0.5 & --- & 0.049$\pm$0.022 & 7.2$\pm$0.8 \\
  & --- & --- & --- & --- & --- & --- & --- & --- & 0.949$\pm$0.014 & 17.5$\pm$0.9 \\
  &  &  &  &  &  &  &  &  &  &  \\
 J1900$-$2600 & --- & --- & --- & 0.135$\pm$0.023 & 7.4$\pm$1.0 & 0.129$\pm$0.011 & 14.1$\pm$1.9 & --- & 0.136$\pm$0.020 & 8.2$\pm$1.2 \\
  & --- & --- & --- & 0.866$\pm$0.023 & 8.0$\pm$0.9 & --- & --- & --- & 0.869$\pm$0.021 & 9.2$\pm$0.9 \\
  &  &  &  &  &  &  &  &  &  &  \\
 J1901$-$0906 & 0.327$\pm$0.011 & 14.7$\pm$0.6 & --- & 0.325$\pm$0.012 & 29.6$\pm$1.2 & 0.329$\pm$0.008 & 17.9$\pm$0.7 & --- & 0.328$\pm$0.010 & 35.2$\pm$1.4 \\
 & 0.138$\pm$0.017 & 8.0$\pm$0.6 & --- & 0.137$\pm$0.011 & 10.0$\pm$0.9 & 0.135$\pm$0.011 & 4.8$\pm$0.6 & --- & 0.138$\pm$0.011 & 11.1$\pm$1.2 \\
 & 0.190$\pm$0.007 & 6.3$\pm$0.5 & --- & 0.189$\pm$0.011 & 6.2$\pm$0.6 & 0.195$\pm$0.010 & 5.4$\pm$0.5 & --- & 0.194$\pm$0.012 & 7.4$\pm$0.9 \\
 & --- & --- & --- & 0.667$\pm$0.011 & 12.5$\pm$0.5 & --- & --- & --- & 0.671$\pm$0.009 & 11.9$\pm$1.0 \\
 & --- & --- & --- & 0.809$\pm$0.007 & 7.1$\pm$0.7 & --- & --- & --- & 0.809$\pm$0.012 & 6.1$\pm$0.8 \\
  &  &  &  &  &  &  &  &  &  &  \\
 J1909+1102  & 0.078$\pm$0.017 & 7.8$\pm$0.9 & --- & 0.073$\pm$0.014 & 15.2$\pm$2.2 & 0.065$\pm$0.029 & 3.9$\pm$0.4 & --- & 0.055$\pm$0.024 & 7.4$\pm$1.0 \\
  & --- & --- & --- & 0.926$\pm$0.025 & 6.9$\pm$0.6 & --- & --- & --- & 0.932$\pm$0.032 & 5.0$\pm$0.7 \\
  &  &  &  &  &  &  &  &  &  &  \\
 J1919+0021  & 0.091$\pm$0.003 & 71.8$\pm$6.2 & --- & 0.091$\pm$0.006 & 86.7$\pm$8.3 & --- & --- & --- & --- & --- \\
  & --- & --- & --- & 0.909$\pm$0.003 & 99.1$\pm$7.6 & --- & --- & --- & --- & --- \\
  &  &  &  &  &  &  &  &  &  &  \\
 J1919+0134  & --- & --- & --- & --- & --- & 0.153$\pm$0.013 & 2.5$\pm$0.3 & --- & 0.152$\pm$0.014 & 18.8$\pm$1.9 \\
  &  &  &  &  &  &  &  &  &  &  \\
 J1921+1948  & 0.168$\pm$0.007 & 4.8$\pm$0.4 & --- & --- & --- & --- & --- & --- & --- & --- \\
  & --- & --- & --- & 0.834$\pm$0.007 & 36.2$\pm$1.5 & --- & --- & --- & --- & --- \\
  & --- & --- & --- & 0.736$\pm$0.017 & 8.6$\pm$0.6 & --- & --- & --- & --- & --- \\
  &  &  &  &  &  &  &  &  &  &  \\
 J1921+2153  & --- & --- & --- & --- & --- & 0.236$\pm$0.013 & 12.8$\pm$0.9 & --- & 0.235$\pm$0.014 & 22.1$\pm$1.4 \\
  & --- & --- & --- & --- & --- & --- & --- & --- & 0.764$\pm$0.012 & 9.8$\pm$0.7 \\
  &  &  &  &  &  &  &  &  &  &  \\
  J1932+1059  & 0.087$\pm$0.010 & 16.3$\pm$2.3 & --- & 0.087$\pm$0.010 & 17.6$\pm$2.6 & --- & --- & --- & --- & --- \\
  & --- & --- & --- & 0.913$\pm$0.010 & 17.3$\pm$1.6 & --- & --- & --- & --- & --- \\
  &  &  &  &  &  &  &  &  &  &  \\
  J1946+1805  & --- & --- & 12.9$\pm$0.8 & --- & --- & 0.036$\pm$0.003 & 101.5$\pm$9.2 & 9.2$\pm$0.4 & 0.037$\pm$0.003 & 127.7$\pm$10.6 \\
  & --- & --- & --- & --- & --- & --- & --- & --- & 0.966$\pm$0.003 & 99.2$\pm$4.2 \\
  & --- & --- & --- & --- & --- & --- & --- & --- & 0.986$\pm$0.003 & 77.4$\pm$3.8 \\
  &  &  &  &  &  &  &  &  &  &  \\
 J2006$-$0807 & 0.027$\pm$0.008 & 10.0$\pm$2.1 & --- & 0.027$\pm$0.008 & 24.9$\pm$5.7 & 0.017$\pm$0.003 & 31.3$\pm$6.7 & --- & 0.017$\pm$0.005 & 61.9$\pm$18.0 \\
  & 0.065$\pm$0.017 & 3.2$\pm$0.4 & --- & 0.066$\pm$0.011 & 15.4$\pm$2.6 & --- & --- & --- & --- & --- \\
  & --- & --- & --- & 0.973$\pm$0.007 & 21.2$\pm$1.9 & --- & --- & --- & 0.982$\pm$0.004 & 58.9$\pm$4.2 \\
  &  &  &  &  &  &  &  &  &  &  \\
 J2046$-$0421 & 0.363$\pm$0.005 & 56.5$\pm$1.5 & 3.37$\pm$0.06 & --- & --- & 0.364$\pm$0.004 & 60.5$\pm$1.8 & 3.15$\pm$0.04 & --- & --- \\
  & --- & --- & --- & 0.637$\pm$0.006 & 54.6$\pm$1.4 & --- & --- & --- & 0.635$\pm$0.005 & 71.0$\pm$1.7 \\
  &  &  &  &  &  &  &  &  &  &  \\
  J2046+1540  & --- & --- & --- & --- & --- & 0.043$\pm$0.011 & 4.1$\pm$0.7 & --- & 0.044$\pm$0.012 & 18.5$\pm$2.8 \\
  & --- & --- & --- & --- & --- & 0.107$\pm$0.022 & 1.1$\pm$0.2 & --- & 0.108$\pm$0.023 & 6.7$\pm$0.8 \\
  &  &  &  &  &  &  &  &  &  &  \\
 J2048$-$1616 & 0.310$\pm$0.003 & 64.9$\pm$9.2 & --- & 0.310$\pm$0.003 & 65.8$\pm$7.0 & --- & --- & --- & --- & --- \\
 & --- & --- & --- & 0.691$\pm$0.002 & 80.7$\pm$10.1 & --- & --- & --- & --- & --- \\
  &  &  &  &  &  &  &  &  &  &  \\
  J2305+3100  & 0.485$\pm$0.010 & 26.0$\pm$0.7 & 4.5$\pm$0.3 & --- & --- & --- & --- & --- & --- & --- \\
  & --- & --- & --- & 0.513$\pm$0.011 & 30.4$\pm$0.9 & --- & --- & --- & --- & --- \\
  &  &  &  &  &  &  &  &  &  &  \\
  J2317+2149  & 0.191$\pm$0.021 & 3.7$\pm$0.8 & 2.2$\pm$0.3 & 0.190$\pm$0.030 & 13.8$\pm$1.9 & 0.194$\pm$0.019 & 2.1$\pm$0.4 & 2.6$\pm$0.2 & 0.193$\pm$0.010 & 15.8$\pm$1.9 \\
  &  &  &  &  &  &  &  &  &  &  \\
 J2330$-$2005 & 0.044$\pm$0.018 & 13.3$\pm$1.4 & --- & 0.041$\pm$0.015 & 16.8$\pm$2.5 & --- & --- & --- & --- & --- \\
  & --- & --- & --- & 0.957$\pm$0.021 & 12.1$\pm$1.1 & --- & --- & --- & --- & --- \\
  &  &  &  &  &  &  &  &  &  &  \\
\hline
\enddata
\label{tab1}
\end{deluxetable}

%% file: table2.tex
\clearpage
\begin{deluxetable}{cccccc}
\tabletypesize{\small}
\tablecolumns{6}
\tablewidth{0pc}
\tablecaption{The table presents the estimated drifting properties along with physical parameters for the pulsars. 
Column 2 and 3 present the period and $\dot{E}$ obtained from the ATNF pulsar database. 
Column 4 presents the type of drifting from three possibilities, ND - negative drifting, PD - positive drifting and AMD - amplitude modulated drifting. 
The pulsar J1034$-$3224 showed a complex phase behaviour which could not be categorized into any of the three forms.
Column 5 presents the peak frequency ($f_p$) of drifting. 
The $f_p$ is the weighted average of the measurements over the two observing frequencies as well as the LRFS and HRFS, see table \ref{peak_mes} for individual measurements. 
In case of multiple peaks for a given pulsar we have only used the most significant peak. 
The two pulsars J1116$-$4122 and J1555$-$3134 had two peaks with equivalent $S$ and both were reported. 
The PD has 0.5~$< f_p <$~1 cycles/$P$ while for ND and AMD 0~$< f_p <$~0.5 cycles/$P$, see text for detail.
Column 6 presents the estimated drift periodicity $P_3$ = 1/$f_p$.}
\tablehead{ 
\multicolumn{1}{c}{PSR} & \multicolumn{1}{c}{$P$} & \multicolumn{1}{c}{$\dot{E}$} & \multicolumn{1}{c}{Type} & \multicolumn{1}{c}{$f_p$} & \multicolumn{1}{c}{$P_3$} \\
\hline
\multicolumn{1}{c}{} & \multicolumn{1}{c}{(s)} & \multicolumn{1}{c}{(10$^{30}$erg~s$^{-1}$)} & \multicolumn{1}{c}{} & \multicolumn{1}{c}{(cycles/$P$)} & \multicolumn{1}{c}{($P$)} \\
\hline
}

\startdata
 J0034$-$0721 & 0.9430 & 19.2 &  ND  & 0.152$\pm$0.012 &  6.6$\pm$0.5  \\
              &        &      &       &                 &               \\
 J0151$-$0635 & 1.4647 & 5.56 &  ND  &0.0695$\pm$0.004 & 14.4$\pm$0.8  \\
              &        &      &       &                 &               \\
 J0152$-$1637 & 0.8327 & 88.8 &  ND  & 0.169$\pm$0.029 &  5.9$\pm$1.0  \\
              &        &      &       &                 &               \\
  J0304+1932  & 1.3876 & 19.1 &  ND  & 0.155$\pm$0.042 &  6.4$\pm$1.7  \\
              &        &      &       &                 &               \\
  J0525+1115  & 0.3544 & 65.3 &  PD  & 0.684$\pm$0.052 &  1.5$\pm$0.1  \\
              &        &      &       &                 &               \\
 J0630$-$2834 & 1.2444 &  146 &  PD  & 0.855$\pm$0.032 & 1.17$\pm$0.04 \\
              &        &      &       &                 &               \\
 J0758$-$1528 & 0.6823 &  201 &  AMD  & 0.044$\pm$0.020 & 22.7$\pm$10.1 \\
              &        &      &       &                 &               \\
 J0820$-$1350 & 1.2381 & 43.8 &  ND  & 0.211$\pm$0.006 &  4.7$\pm$0.1  \\
              &        &      &       &                 &               \\
  J0837+0610  & 1.2738 &  130 &  AMD  & 0.460$\pm$0.005 & 2.17$\pm$0.03 \\
              &        &      &       &                 &               \\
 J0846$-$3533 & 1.1161 & 45.5 &  ND  & 0.492$\pm$0.005 & 2.03$\pm$0.02 \\
              &        &      &       &                 &               \\
 J0944$-$1354 & 0.5703 & 9.63 &  ND  & 0.156$\pm$0.008 &  6.4$\pm$0.3  \\
              &        &      &       &                 &               \\
 J0959$-$4809 & 0.6701 & 10.8 &  ND  & 0.179$\pm$0.041 &  5.6$\pm$1.3  \\
              &        &      &       &                 &               \\
 J1034$-$3224 & 1.1506 & 5.97 &  ---  & 0.139$\pm$0.009 &  7.2$\pm$0.5  \\
              &        &      &       &                 &               \\
 J1041$-$1942 & 1.3864 & 14.0 &  ND  & 0.233$\pm$0.024 &  4.3$\pm$0.4  \\
              &        &      &       &                 &               \\
 J1116$-$4122 & 0.9432 &  374 &  AMD  & 0.028$\pm$0.011 & 36.4$\pm$13.9 \\
              &        &      &       & 0.050$\pm$0.011 & 20.0$\pm$4.3  \\
              &        &      &       &                 &               \\
  J1239+2453  & 1.3824 & 14.3 &  AMD  & 0.361$\pm$0.014 &  2.8$\pm$0.1  \\
              &        &      &       &                 &               \\
 J1328$-$4921 & 1.4787 & 7.45 &  AMD  & 0.294$\pm$0.018 &  3.4$\pm$0.2  \\
              &        &      &       &                 &               \\
 J1418$-$3921 & 1.0968 & 26.6 &  PD  & 0.598$\pm$0.005 & 1.67$\pm$0.01 \\
              &        &      &       &                 &               \\
 J1527$-$3931 & 2.4176 & 53.3 &  PD  & 0.975$\pm$0.004 &1.026$\pm$0.004\\
              &        &      &       &                 &               \\
 J1555$-$3134 & 0.5181 & 17.7 &  ND  & 0.097$\pm$0.008 & 10.3$\pm$0.9  \\
              &        &      &       & 0.054$\pm$0.011 & 18.5$\pm$3.9  \\
              &        &      &       &                 &               \\
 J1603$-$2531 & 0.2831 &2.77$\times$10$^3$&  AMD  &0.0145$\pm$0.006 & 69.0$\pm$28.5 \\
              &        &      &       &                 &               \\
 J1604$-$4909 & 0.3274 &1.15$\times$10$^3$&  AMD  & 0.020$\pm$0.010 & 50.0$\pm$26.1 \\
              &        &      &       &                 &               \\
 J1625$-$4048 & 2.3553 & 1.34 &  AMD  &0.0165$\pm$0.010 & 60.6$\pm$36.7 \\
              &        &      &       &                 &               \\
 J1645$-$0317 & 0.3877 &1.21$\times$10$^3$&  AMD  & 0.080$\pm$0.028 & 12.5$\pm$4.4  \\
              &        &      &       &                 &               \\
 J1700$-$3312 & 1.3583 & 74.2 &  PD  & 0.553$\pm$0.022 & 1.81$\pm$0.07 \\
              &        &      &       &                 &               \\
 J1703$-$3241 & 1.2118 & 14.6 &  PD  & 0.214$\pm$0.025 &  4.7$\pm$0.5  \\
              &        &      &       &                 &               \\
 J1720$-$2933 & 0.6204 &  123 &  ND  &0.4085$\pm$0.003 & 2.45$\pm$0.02 \\
              &        &      &       &                 &               \\
 J1722$-$3207 & 0.4772 &  235 &  AMD  & 0.044$\pm$0.018 & 22.7$\pm$9.4  \\
              &        &      &       &                 &               \\
 J1733$-$2228 & 0.8717 & 2.55 &  AMD  & 0.044$\pm$0.026 & 22.9$\pm$13.7 \\
              &        &      &       &                 &               \\
 J1733$-$3716 & 0.3376 &1.54$\times$10$^4$&  AMD  & 0.014$\pm$0.004 & 71.4$\pm$20.4 \\
              &        &      &       &                 &               \\
 J1735$-$0724 & 0.4193 &  650 &  AMD  & 0.051$\pm$0.019 & 19.7$\pm$7.5  \\
              &        &      &       &                 &               \\
  J1740+1311  & 0.8030 &  111 &  AMD  & 0.113$\pm$0.029 &  8.8$\pm$2.3  \\
              &        &      &       &                 &               \\
 J1741$-$3927 & 0.5122 &  567 &  AMD  & 0.106$\pm$0.047 &  9.5$\pm$4.2  \\
              &        &      &       &                 &               \\
 J1741$-$0840 & 2.0431 & 10.5 &  PD  & 0.782$\pm$0.027 & 1.28$\pm$0.04 \\
              &        &      &       &                 &               \\
 J1748$-$1300 & 0.3941 &  782 &  AMD  & 0.150$\pm$0.014 &  6.7$\pm$0.6  \\
              &        &      &       &                 &               \\
 J1801$-$0357 & 0.9215 &  167 &  AMD  & 0.012$\pm$0.006 & 81.4$\pm$41.9 \\
              &        &      &       &                 &               \\
 J1801$-$2920 & 1.0819 &  103 &  AMD  & 0.404$\pm$0.013 & 2.48$\pm$0.08 \\
              &        &      &       &                 &               \\
 J1816$-$2650 & 0.5929 & 12.6 &  ND  & 0.244$\pm$0.013 &  4.1$\pm$0.2  \\
              &        &      &       &                 &               \\
 J1822$-$2256 & 1.8743 & 8.12 &  ND  & 0.058$\pm$0.008 & 17.2$\pm$2.3  \\
              &        &      &       &                 &               \\
 J1842$-$0359 & 1.8399 & 3.22 &  AMD  & 0.081$\pm$0.003 & 12.4$\pm$0.5  \\
              &        &      &       &                 &               \\
 J1848$-$0123 & 0.6594 &  723 &  AMD  & 0.050$\pm$0.020 & 19.9$\pm$7.8  \\
              &        &      &       &                 &               \\
 J1900$-$2600 & 0.6122 & 35.2 &  AMD  & 0.131$\pm$0.015 &  7.6$\pm$0.8  \\
              &        &      &       &                 &               \\
 J1901$-$0906 & 1.7819 & 11.4 &  ND  & 0.328$\pm$0.010 & 3.05$\pm$0.09 \\
              &        &      &       &                 &               \\
  J1909+1102  & 0.2836 &4.57$\times$10$^3$&  AMD  & 0.071$\pm$0.018 & 14.1$\pm$3.6 \\
              &        &      &       &                 &               \\
  J1919+0021  & 1.2723 &  147 &  AMD  & 0.091$\pm$0.004 & 11.0$\pm$0.4  \\
              &        &      &       &                 &               \\
  J1919+0134  & 1.6040 & 5.63 &  ND  & 0.153$\pm$0.013 &  6.6$\pm$0.6  \\
              &        &      &       &                 &               \\
  J1921+1948  & 0.8210 & 63.9 &  PD  & 0.834$\pm$0.007 & 1.20$\pm$0.01 \\
              &        &      &       &                 &               \\
  J1921+2153  & 1.3373 & 22.3 &  ND  & 0.236$\pm$0.013 &  4.2$\pm$0.2  \\
              &        &      &       &                 &               \\
  J1932+1059  & 0.2265 &3.93$\times$10$^3$&  AMD  & 0.087$\pm$0.010 & 11.5$\pm$1.3 \\
              &        &      &       &                 &               \\
  J1946+1805  & 0.4406 & 11.1 &  AMD  &0.0365$\pm$0.003 & 27.4$\pm$2.3  \\
              &        &      &       &                 &               \\
 J2006$-$0807 & 0.5809 & 9.27 &  AMD  & 0.019$\pm$0.004 & 53.4$\pm$12.3 \\
              &        &      &       &                 &               \\
 J2046$-$0421 & 1.5469 & 15.7 &  PD  & 0.636$\pm$0.005 & 1.57$\pm$0.01 \\
              &        &      &       &                 &               \\
  J2046+1540  & 1.1383 & 4.88 &  ND  & 0.043$\pm$0.011 & 23.0$\pm$6.1  \\
              &        &      &       &                 &               \\
 J2048$-$1616 & 1.9616 & 57.3 &  AMD  & 0.310$\pm$0.003 & 3.23$\pm$0.03 \\
              &        &      &       &                 &               \\
  J2305+3100  & 1.5759 & 29.2 &  PD  & 0.513$\pm$0.011 & 1.95$\pm$0.04 \\
              &        &      &       &                 &               \\
  J2317+2149  & 1.4447 & 13.7 &  ND  & 0.193$\pm$0.014 &  5.2$\pm$0.4  \\
              &        &      &       &                 &               \\
 J2330$-$2005 & 1.6436 & 41.2 &  AMD  & 0.042$\pm$0.016 & 23.7$\pm$9.1  \\
              &        &      &       &                 &               \\
\hline
\enddata
\label{tab2}
\end{deluxetable}